\documentclass[twocolumn]{aa}
\usepackage[latin1]{inputenc}
\usepackage[T1]{fontenc}
\usepackage{lmodern}
\usepackage[english]{babel}
\usepackage{epsfig}
\usepackage{lscape}
\usepackage{float}
\usepackage{amsmath}
\usepackage{amssymb}
\usepackage{wasysym}
\usepackage{multirow}
\usepackage{graphicx}

\begin{document}

%\begin{multicols}{2}

\title{Titan's rotation}
\subtitle{A 3-dimensional theory}

\author{B.~Noyelles\inst{1,2} \and A.~Lemaître\inst{1} \and A.~Vienne\inst{2,3}}

\institute{University of Namur - Department of Mathematics - Rempart de la Vierge 8 - 5000 Namur - Belgium
\and IMCCE, Paris Observatory, UPMC, Univ. Lille 1, CNRS UMR 8028 - 77 avenue Denfert Rochereau - 75014 Paris - France
\and LAL / University of Lille - 1 impasse de l'Observatoire - 59000 Lille - France}

\offprints{B.~Noyelles, e-mail: noyelles@imcce.fr}

\date{Received / Accepted}

\abstract
{}
{\par We study the forced rotation of Titan seen as a rigid body at the equilibrium Cassini state, involving the spin-orbit synchronization.}
{\par We used both the analytical and the numerical ways. We analytically determined the equilibrium positions and the frequencies of the 3 free librations around it, while a numerical integration associated to frequency analysis gave us a more synthetic, complete theory, where the free solution split from the forced one. }
{\par We find a mean obliquity of 2.2 arcmin and the fundamental frequencies of the free librations of about 2.0977, 167.4883, and 306.3360 years. Moreover, we bring out the main role played by Titan's inclination on its rotation, and we suspect a likely resonance involving Titan's wobble.}
{}


\maketitle

\titlerunning{Titan's rotation}
\authorrunning{Noyelles et al.}

%\begin{multicols}{2}
\section{Introduction}

\par Since the terrestrial observations of Lemmon et al. (\cite{Lemmon93}), the rotation of Titan, Saturn's main satellite, has been assumed to be synchronous or nearly synchronous. This has been confirmed by Lemmon et al. (\cite{Lemmon95}) and by Richardson et al. (\cite{Richardson04}) with the help of Voyager I images. In this last work, Titan's rotation period is estimated at $15.9458 \pm 0.0016$ days, whereas its orbital period is  $15.945421 \pm 0.000005$ days.

\par The spin-orbit synchronization of a natural satellite is very common in the solar system (such as for the Moon and the Galilean satellites of Jupiter) and is known as a Cassini state. This is an equilibrium state that has probably been reached after a deceleration of the spin of the involved body under dissipative effects, like tides.

\par Recently, Henrard and Schwanen (\cite{Henrard04}) have given a 3-dimensional elaborated analytical model of the forced rotation of synchronous triaxial bodies, after studying the librations around the Cassini state. This model has been successfully applied by Henrard on the Galilean satellites Io (\cite{Henrard05i}) and Europa (\cite{Henrard05c}), seen as rigid bodies. Such studies require knowing some parameters of the gravitational field of the involved bodies, which cannot be considered as spheres. Another analytical study has been performed for Mercury by D'Hoedt and Lemaître (\cite{DHoedt04}), for the case of a $3:2$ spin-orbit resonance.

\par Since the first fly-bys of Titan by the Cassini spacecraft, we have a first estimation of the useful parameters, more particularly Titan's $J_2$ and $C_{22}$ (Tortora et al. \cite{Tortora06}), so a similar study of Titan's rotation can be made. In this paper, we propose a study of Titan's forced rotation, where Titan is seen as a rigid body. The originality of this study over Henrard's previous studies is that we use both the analytical and the numerical tools and compare our results.

\section{Expressing the problem}

\par Titan is here considered as a triaxial rigid body whose principal moments of inertia are written respectively as $A$, $B$, and $C$, with $A \leq B \leq C$.

\subsection{The variables}

\par Our variables and equations have already been used in previous studies; see for instance Henrard and Schwanen (\cite{Henrard04}) for the general case of synchronous satellites, Henrard (\cite{Henrard05i}) for Io, and Henrard (\cite{Henrard05c}) for Europa.

\par We consider 3 reference frames : the first $(\vec{e_1},\vec{e_2},\vec{e_3})$ is centered on Titan's mass barycenter and is in translation with the inertial reference frame used to describe the orbital motion of the Saturnian satellites in the TASS1.6 theory (see Vienne \& Duriez \cite{Vienne95}). This is a cartesian coordinate system whose origin is the center of Saturn, and it refers to the equatorial plane of Saturn and the node of this plane with the ecliptic at J2000. The second frame $(\vec{n_1},\vec{n_2},\vec{n_3})$ is linked to Titan's angular momentum, and the third one $(\vec{f_1},\vec{f_2},\vec{f_3})$ is rigidly linked to Titan. In this last frame, Titan's matrix of inertia is written as 

\begin{equation}
I=\left(\begin{array}{ccc}
A & 0 & 0 \\
0 & B & 0 \\
0 & 0 & C
\end{array}\right).
\label{equ:inertie}
\end{equation}

\par
\begin{figure*}[ht]
\centering
\includegraphics[width=13.5cm]{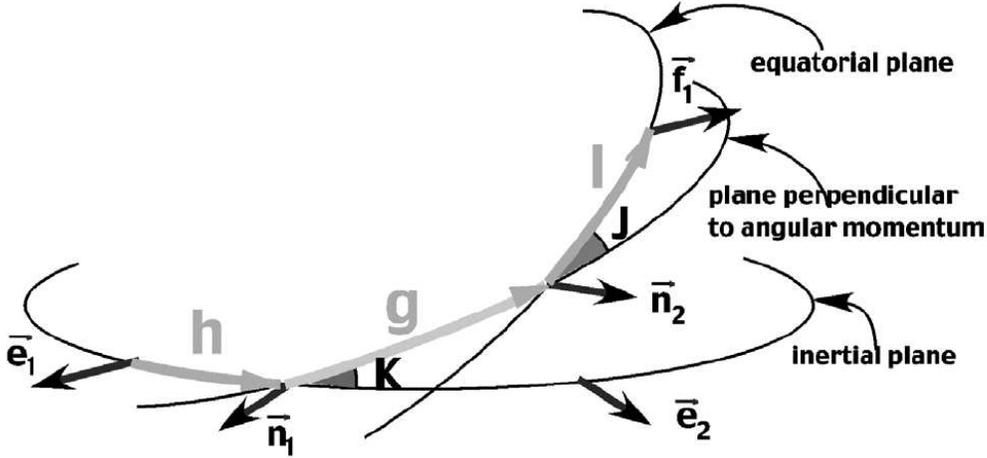}
\caption{The Andoyer variables (reproduced from Henrard \cite{Henrard05i}).}
\label{fig:angles}
\end{figure*}

 We first use Andoyer's variables (see Andoyer \cite{Andoyer26} and Deprit \cite{Deprit67}), which are based on two linked sets of Euler's angles. The first set $(h,K,g)$ locates the position of the angular momentum in the first frame $(\vec{e_1},\vec{e_2},\vec{e_3})$, while the second $(g,J,l)$ locates the body frame $(\vec{f_1},\vec{f_2},\vec{f_3})$ in the second frame tied to the angular momentum (see Fig. \ref{fig:angles}).

\par The canonical set of Andoyer's variables consists of the three angular variables $l,g,h$ and their conjugated momenta defined by the norm $G$ of the angular momentum and two of its projections: 

\begin{center}
$\begin{array}{lll}
l & \hspace{3cm} & L=G\cos J \\
g & \hspace{3cm} & G \\
h & \hspace{3cm} & H=G\cos K
\end{array}$
\end{center}

\par Unfortunately, these variables present two singularities: when $J=0$ (i.e., the angular momentum is colinear to $\vec{f_3}$, there is no wobble), $l$ and $g$ are undefined, and when $K=0$ (i.e., when Titan's principal axis of inertia is perpendicular to its orbital plane), $h$ and $g$ are undefined. That is why we use the modified Andoyer's variables:

\begin{center}
$\begin{array}{lll}
p=l+g+h & \hspace{2cm} & P=\frac{G}{nC} \\
r=-h & \hspace{2cm} & R=\frac{G-H}{nC}=P(1-\cos K) \\
 & \hspace{2cm} & =2P\sin^2\frac{K}{2} \\
\xi_q=\sqrt{\frac{2Q}{nC}}\sin q & \hspace{2cm} & \eta_q=\sqrt{\frac{2Q}{nC}}\cos q \\
\end{array}$ \\
\end{center}
where $n$ is Titan's mean orbital motion , $q=-l$, and $Q=G-L=G(1-\cos J)=2G\sin^2\frac{J}{2}$. With these new variables, the singularity on $l$ has been dropped.

\subsection{The free rotation}

\par To describe the dynamics of the system, we should consider the free rotation and the perturbations by other bodies. The Hamiltonian of the free body rotation is also the kinetic energy of the rotation $T=\frac{1}{2}(\vec{\omega}|\vec{G})$ where $\vec{\omega}$ is the instantaneous rotation vector, $\vec{G}$ the angular momentum vector with respect to the center of mass, and $(\vec{\omega}|\vec{G})$ the scalar product of the vector $\vec{\omega}$ and $\vec{G}$, where $\vec{\omega}$ and $\vec{G}$ are respectively defined as

\begin{equation}
\vec{\omega}=\omega_1 \vec{f_1}+\omega_2 \vec{f_2}+\omega_3 \vec{f_3}
\label{equ:omega0}
\end{equation}
and

\begin{equation}
\vec{G}=A\omega_1 \vec{f_1}+B\omega_2 \vec{f_2}+C\omega_3 \vec{f_3}.
\label{equ:G0}
\end{equation}

We also deduce from the definitions of the angles $l$ and $J$ (the wobble):

\begin{equation}
\vec{G}=G\sin J\sin l \vec{f_1}+G\sin J\cos l\vec{f_2}+G\cos J \vec{f_3},
\label{equ:G}
\end{equation}
from which we can easily deduce

\begin{equation}
\vec{\omega}=\frac{G}{A}\sin J\sin l \vec{f_1}+\frac{G}{B}\sin J\cos l\vec{f_2}+\frac{G}{C}\cos J \vec{f_3}
\label{equ:omega}
\end{equation}
and consequently

\begin{equation}
\frac{1}{2}(\vec{\omega}|\vec{G})=\frac{G^2-L^2}{2}\Bigg[\frac{\sin^2 l}{A}+\frac{\cos^2 l}{B}\Bigg]+\frac{L^2}{2C}.
\label{equ:Tando}
\end{equation}
As a result, the Hamiltonian of the free rotation in the modified Andoyer's variables is 

\begin{equation}
T=\frac{nP^2}{2}+\frac{nP}{8}\Big[4-\xi_q^2-\eta_q^2\Big]\Bigg[\frac{\gamma_1+\gamma_2}{1-\gamma_1-\gamma_2}\xi_q^2
+\frac{\gamma_1-\gamma_2}{1-\gamma_1+\gamma_2}\eta_q^2\Bigg]
\label{equ:Tandomod}
\end{equation}
with 

\begin{equation}
\gamma_1=\frac{2C-A-B}{2C}=J_2\frac{MR^2}{C}
\label{equ:gamma1}
\end{equation}
and

\begin{equation}
\gamma_2=\frac{B-A}{2C}=2C_{22}\frac{MR^2}{C}.
\label{equ:gamma2}
\end{equation}

\subsection{Perturbation by Saturn}

\par Considering the parent body Saturn as a point mass $M_{\saturn}$, the gravitational potential of the perturbation can be written as

\begin{equation}
V=-\mathcal{G}M_{\saturn}\int\int\int_W \frac{\rho dW}{d'}
\label{equ:Vsatu}
\end{equation}
where $\rho$ is the density inside the volume $W$ of the body and $d'$ the distance between the point mass and a volume element inside the body. Using the usual expansion of the potential in spherical harmonics (see for instance Bertotti and Farinella \cite{Bertotti90}), we find

\begin{equation}
\begin{split}
V=-\frac{\mathcal{G}M_{\saturn}}{d}\bigg(1+\sum_{n\geq 1}\frac{1}{d^n}\sum_{m=0}^nP_n^m(\sin\phi)\times \\
\big[C_n^m\cos m\psi + S_n^m\sin m\psi \big] \bigg)
\end{split}
\label{equ:Vsatudev}
\end{equation}
where $\psi$ and $\phi$ are respectively the longitude and the latitude of Saturn's barycenter of mass in Titan's frame, and $d$ the distance between this Saturn's barycenter of mass and the origin of the frame (Titan's barycenter of mass). If we limit the expansion of (\ref{equ:Vsatudev}) to the second order terms and drop the term $\frac{\mathcal{G}M_{\saturn}}{d}$, which does not produce any effect on the rotation, we have

\begin{equation}
V=-\frac{3\mathcal{G}M_{\saturn}}{2d^3}MR^2\Big[J_2(x^2+y^2)+2c_{22}(x^2-y^2)\Big]
\label{equ:Vsatudevxy}
\end{equation}
where $x$, $y$, and $z$ are the coordinates of Saturn's center of mass in Titan's frame $(\vec{f_1},\vec{f_2},\vec{f_3})$ (so we have $d=\sqrt{x^2+y^2+z^2}$). Here, $d$ depends on the time since Titan's motion around Saturn is not circular, but we can introduce $d_0$, the mean value of $d$, since $a_0$ is Saturn's mean semimajor axis and $e_0$ its mean eccentricity. (They correspond respectively to Titan's mean semimajor axis and mean eccentricity in a Saturnian frame.)

\par We use the formula

\begin{equation}
d_0=a\bigg(1+\frac{e^2}{2}\bigg)
\label{equ:d0}
\end{equation}
coming from the development of $\frac{r}{a}$ (see Brouwer \& Clemence \cite{Brouwer61}):

\begin{equation}
\begin{split}
\frac{r}{a}=1+\frac{e^2}{2}+\bigg(-e+\frac{3}{8}e^3\bigg)\cos\mathcal{M} \\
-\frac{e^2}{2}\cos 2\mathcal{M} -\frac{3}{8} e^3 \cos 3\mathcal{M}
+O(e^4)
\end{split}
\label{equ:brouwer}
\end{equation}
from which the mean anomaly $\mathcal{M}$ disappears after averaging. The perturbing potential $V$ now reads

\begin{equation}
V=-\frac{3}{2}\frac{\mathcal{G}M_{\saturn}}{d_0^3}\bigg(\frac{d_0}{d}\bigg)^3MR^2\Big[J_2(x^2+y^2)+2c_{22}(x^2-y^2)\Big]
\label{equ:Vsatudevxyd0}
\end{equation}
and we set $n^{*2}=\frac{\mathcal{G}M_{\saturn}}{d_0^3}$, so that we can write

\begin{equation}
\frac{V}{nC}=n\bigg(\frac{d}{d_0}\bigg)^3\Big[\delta_1(x^2+y^2)+\delta_2(x^2-y^2)\Big]
\label{equ:VsnC}
\end{equation}
with

\begin{equation}
\delta_1=-\frac{3}{2}\bigg(\frac{n^*}{n}\bigg)^2\gamma_1
\label{equ:delta1}
\end{equation}
and

\begin{equation}
\delta_2=-\frac{3}{2}\bigg(\frac{n^*}{n}\bigg)^2\gamma_2
\label{equ:delta2}
\end{equation}
where $M$ and $R$ are respectively Titan's mass and radius.

\par As Henrard (\cite{Henrard05c}) did for Jupiter, we also take Saturn's oblateness into account. The perturbing potential due to Saturn's oblateness reads

\begin{equation}
V_o=\delta_sCn^2\bigg(\frac{d_0}{d}\bigg)^5\big[\delta_1(x^2+y^2)+\delta_2(x^2-y^2)\big]
\label{equ:vo}
\end{equation}
with
\begin{equation}
\delta_s=\frac{5}{2}J_{2{\saturn}}\Bigg(\frac{R_{\saturn}}{d_0}\Bigg)^2
\label{equ:deltas}
\end{equation}
where $R_{\saturn}$ is Saturn's radius, and $J_{2{\saturn}}$ its $J_2$.

\par Finally, the Hamiltonian of the problem reads 

\begin{equation}
\begin{split}
\mathcal{H}=\frac{nP^2}{2} \\
+\frac{nP}{8}\Big[4-\xi_q^2-\eta_q^2\Big]\bigg[\frac{\gamma_1+\gamma_2}{1-\gamma_1-\gamma_2}\xi_q^2
+\frac{\gamma_1-\gamma_2}{1-\gamma_1+\gamma_2}\eta_q^2\bigg] \\
+n\bigg(\frac{d_0}{d}\bigg)^3\Bigg(1+
\delta_s\bigg(\frac{d_0}{d}\bigg)^2\Bigg)\big[\delta_1(x^2+y^2)+\delta_2(x^2-y^2)\big].
\end{split}
\label{equ:hamilton}
\end{equation}

\section{Analytical study}

\par We intend to use the Hamiltonian (\ref{equ:hamilton}) to analytically determine the equilibrium position of Titan in the Cassini state related to the spin-orbit synchronization and the 3 frequencies of the free librations around this equilibrium, using the method explained in (Henrard \& Schwanen \cite{Henrard04}). For this analytical study, we consider that Titan has a circular orbit around Saturn, whose inclination on Saturn's equatorial plane is given by only one periodic term extracted from TASS1.6 ephemerides (Vienne \& Duriez \cite{Vienne95}). This implies that the ascending node of Titan oscillates around a fixed value, so it cannot disappear after averaging the equations. That is why the analytical solutions of (Henrard and Schwanen \cite{Henrard04}) cannot be used directly, and we first must check that they become unchanged without averaging the ascending node. The true orbital eccentricity of Titan is about $0.0289$, but the opportunity to neglect it will be discussed later, after comparison with the numerical study.

\par This way, the vector locating Saturn's barycenter is colinear to

\begin{equation}
x_i\vec{e_1}+y_i\vec{e_2}+z_i\vec{e_3}
\label{equ:vecei}
\end{equation}
with
\begin{equation}
x_i=-\Big(\cos\ascnode_6\cos(\lambda_6-\ascnode_6)-\cos I_6\sin\ascnode_6\sin(\lambda_6-\ascnode_6)\Big)
\label{equ:xi6}
\end{equation}

\begin{equation}
y_i=-\Big(\sin\ascnode_6\cos(\lambda_6-\ascnode_6)+\cos I_6\cos\ascnode_6\sin(\lambda_6-\ascnode_6)\Big)
\label{equ:yi6}
\end{equation}
and

\begin{equation}
z_i=-\sin I_6\sin \Big(\lambda_6-\ascnode_6\Big)
\label{equ:zi6}
\end{equation}
where $I_6$, $\ascnode_6$ and $\lambda_6$ are respectively Titan's mean inclination, argument of the node and mean longitude in the inertial frame of the ephemerides. (The subscript $6$ refers to the fact that Titan is Saturn's sixth satellite.)

\par To obtain the coordinates $x$, $y$, and $z$ of Saturn is the reference frame bound to Titan $(\vec{f_1},\vec{f_2},\vec{f_3})$, 5 rotations are to be performed:

\begin{equation}
\left(\begin{array}{c}
x \\
y \\
z
\end{array}\right)
=R_3(-l)R_1(-J)R_3(-g)R_1(-K)R_3(-h)\left(\begin{array}{c}
x_i \\
y_i \\
z_i
\end{array}\right)
\label{equ:passage}
\end{equation}
with 

\begin{equation}
R_3(\phi)=\left(\begin{array}{ccc}
\cos\phi & -\sin\phi & 0 \\
\sin\phi & \cos\phi & 0 \\
0 & 0 & 1
\end{array}\right)
\label{equ:r3}
\end{equation}
and

\begin{equation}
R_1(\phi)=\left(\begin{array}{ccc}
1 & 0 & 0 \\
0 & \cos\phi & -\sin\phi \\
0 & \sin\phi & \cos\phi 
\end{array}\right).
\label{equ:r1}
\end{equation}

\par Table \ref{tab:data} gives the values of the physical and dynamical parameter that we use, and Table \ref{tab:coef} gathers the computed values of the corresponding parameters used in the Hamiltonian (\ref{equ:hamilton}).

\begin{table*}[ht]
\centering
\caption{Physical and dynamical parameters.}
\begin{tabular}{l|ll}
Parameters & Values & References \\
\hline
$n$ & $143.9240478491399 rad.y^{-1}$ & TASS1.6  \cite{Vienne95} \\
$e$ & $0.0289$ & TASS1.6 \cite{Vienne95} \\
$\gamma=\sin \frac{I_6}{2}$ & $5.6024\times10^{-3}$ & TASS1.6 \cite{Vienne95} \\
$R_{\saturn}$ & $58232$ km & IAU 2000 \cite{Seidelmann02} \\
$J_{2\saturn}$ & $1.6298\times10^{-2}$ & Pioneer \& Voyager \cite{Campbell89} \\
$M$ & $2.36638\times10^{-4}M_{\saturn}$ & Pioneer \& Voyager \cite{Campbell89} \\
$R$ & $2575$ km & IAU 2000 \cite{Seidelmann02} \\
$\mathcal{G}M_{\saturn}$ & $3.77747586645\times10^{22}.km^3.y^{-2}$ & Pioneer, Voyager + IERS 2003 \\
$J_{2}$ & $(3.15\pm0.32)\times10^{-5}$ & Cassini \cite{Tortora06} \\
$c_{22}$ & $(1.1235\pm0.0061)\times10^{-5}$ & Cassini \cite{Tortora06} \\
$\frac{C}{MR^2}$ & $0.31$ & $(\ldots)$ \\
\hline
\end{tabular}
\begin{note}
The mean values of Titan's mean motion $n$, eccentricity $e$ and inclination $\gamma$ come from TASS1.6 theory (Vienne \& Duriez \cite{Vienne95}), the radii come from the IAU 2000 recommendations (Seidelmann et al. \cite{Seidelmann02}), Titan's mass $M$ and Saturn's $J_2$ come from the Pioneer and Voyager space missions (Campbell \& Anderson \cite{Campbell89}). These two values are those used in TASS1.6 theory, we choose to keep them in order to remain coherent. The mass of Saturn has been derived from the fly-bys of the Pioneer and Voyager space missions, but the published value is given in solar masses. That is why we also indicate IERS 2003 as a reference, which gives us the solar mass. Titan's J2 and C22 come from the fly-by T11 of the Cassini space mission (Tortora et al. \cite{Tortora06}), but unfortunately no value for $\frac{C}{MR^2}$ is available yet. We can only hypothesize that it should be included between $0.3$ and $0.4$, as the case for the Galilean satellites of Jupiter.
\end{note}
\label{tab:data}
\end{table*}

\begin{table}[ht]
\centering
\caption{Values used in the Hamiltonian (\ref{equ:hamilton}) that have been computed from the physical and orbital parameters given Table \ref{tab:data}.}
\begin{tabular}{l|l}
Parameter & Numerical value \\
\hline
$d_0$ & $1222345.284$ km  \\
$n^*$ & $143.8339397847 rad.y^{-1}$ \\
$\gamma_1$ & $1.016129\times10^{-4}$ \\
$\gamma_2$ & $7.248387\times10^{-5}$ \\
$\delta_1$ & $-1.522286\times10^{-4}$ \\
$\delta_2$ & $-1.085897\times10^{-4}$ \\
$\delta_s$ & $9.247193\times10^{-5}$  \\
\hline
\end{tabular}
\label{tab:coef}
\end{table}

\subsection{Equilibrium}

\par We consider here that the system is exactly at the Cassini state. This implies that:

\begin{itemize}

\item The axis of least inertia, $\vec{f_1}$, points to the center of mass of Saturn, so we have $p-\lambda_{\saturn}=0$, $\lambda_{\saturn}$ as the mean longitude of Saturn in the frame $(\vec{f_1},\vec{f_2},\vec{f_3})$.

\item The ascending node of the frame $(\vec{n_1},\vec{n_2},\vec{n_3})$ (associated to the angular momentum) in the inertial frame has the same precession rate as the ascending node of Saturn in the same inertial frame, i.e. $r+\ascnode_{\saturn}=0$, $\ascnode_{\saturn}$ is the argument of the ascending node of Saturn.

\item There is no wobble, so the angular momentum is colinear with Titan's axis of highest inertia $\vec{f_3}$. This implies $J=0$, so $\xi_q=0$ and $\eta_q=0$.

\end{itemize}

\par We have

\begin{equation}
\lambda_{\saturn}=\lambda_{6}-\pi
\label{equ:lambda}
\end{equation}
and

\begin{equation}
\ascnode_{\saturn}=\ascnode_6,
\label{equ:noeud}
\end{equation}
so it is convenient to introduce this new set of canonical variables:

\begin{center}
$\begin{array}{lll}
\sigma=p-\lambda_6+\pi & \hspace{2.8cm} & P \\
\rho=r+\ascnode_6 & \hspace{2.8cm} & R \\
\end{array}$
\end{center}
where $\sigma$ represents the angle between the axis of least inertia of Titan $\vec{f_1}$ and the direction Saturn-Titan, and $\rho$ is the difference between the two ascending nodes. At the exact equilibrium, these two angles should be zero.

\par This way, and also assuming $d \approx d_0$ (i.e., neglecting Titan's orbital eccentricity), the Hamiltonian (\ref{equ:hamilton}) becomes

\begin{equation}
\begin{split}
\mathcal{H}=\frac{nP^2}{2}-nP+\dot{\ascnode}R \\
+n\delta_1(1+\delta_s)[a_1\sin^2K+a_2\sin K\cos K\cos\rho  \\
+a_3\cos2\rho(1-\cos2K)] \\
+n\delta_2(1+\delta_s)[b_1(1+\cos K)^2\cos2\sigma \\
+b_2\sin K(1+\cos K)\cos(2\sigma+\rho) \\
+b_3\sin^2K\cos(2\sigma+2\rho) \\
+b_4\sin K(1-\cos K)\cos(2\sigma+3\rho) \\
+b_5(1-\cos K)^2\cos(2\sigma+4\rho)],
\end{split}
\label{equ:hamsimp}
\end{equation}
with the mean longitude disappearing after averaging, except of course in the $p$ variable. The term $-nP+\dot{\ascnode}R$ has to be added because the canonical transformation we use is time-dependent. The Hamiltonian (\ref{equ:hamsimp}) has been computed with Maple software, and the analytical expressions of the coefficients $a_i$ and $b_i$ are the same as in Henrard and Schwanen (\cite{Henrard04}). This means that, assuming that the ascending node of the orbit of the considered body circulates or not does not change the expressions of $a_i$ and $b_i$, the formulae given in Henrard and Schwanen (\cite{Henrard04}) can be applied to bodies whose node does not circulate, e.g. J-4 Callisto, S-6 Titan or S-8 Iapetus. The analytical expressions of these coefficients are recalled in App.\ref{sec:ab}, while Table \ref{tab:ab} gives their numerical values in our context.

\begin{table}
\centering
\caption{Numerical values of $a_i$ and $b_i$.}
\begin{tabular}{l|l}
Parameter & Numerical Value \\
\hline
$a_1$ & $-4.9990584229813\times10^{-1}$ \\
$a_2$ & $1.12039208002146\times10^{-2}$ \\
$a_3$ & $1.56929503117011\times10^{-5}$ \\
$b_1$ & $2.49984306803404\times10^{-1}$ \\
$b_2$ & $5.60213623923651\times10^{-3}$ \\
$b_3$ & $4.70788509351034\times10^{-5}$ \\
$b_4$ & $1.75839129195533\times10^{-7}$ \\
$b_5$ & $2.46284149427823\times10^{-10}$ \\
\hline
\end{tabular}
\label{tab:ab}
\end{table}

\par At the exact equilibrium, we have $\sigma=0$, $\rho=0$, $\frac{d\sigma}{dt}=\frac{\partial \mathcal{H}}{\partial P}=0$, and $\frac{d\rho}{dt}=\frac{\partial \mathcal{H}}{\partial R}=0$. These two last equations give

\begin{equation}
E_1(P,K)=n\Big[P-1+\big(1+\delta_s\big)\Delta\frac{\cos K-1}{P\sin K}\Big]=0
\label{equ:dhsdp2}
\end{equation}
and

\begin{equation}
E_2(P,K)=\dot{\ascnode}_6+\big(1+\delta_s\big)\frac{n\Delta}{P\sin K}=0
\label{equ:dhsdr2}
\end{equation}
with

\begin{equation}
\begin{split}
\Delta=\delta_1\big[a_1\sin2K+a_2\cos2K+2a_3\sin2K\big] \\
+\delta_2\big[-2b_1\sin K(1+\cos K)+b_2(\cos K+\cos2K) \\
+b_3\sin2K+b_4(\cos K-\cos2K) \\
+2b_5\sin K(1-\cos K)\big].
\end{split}
\label{equ:gd}
\end{equation}

\par Since Titan's ascending node oscillates around a fixed value, we have $\dot{\ascnode}_6=0$. A numerical resolution of (\ref{equ:dhsdp2}) and (\ref{equ:dhsdr2}) gives

\begin{equation}
\begin{split}
K^*=1.1204858615 \times10^{-2} rad \\
= 2311.168 arcsec = 38'31.168" 
\end{split}
\label{equ:Kstar}
\end{equation}

\begin{equation}
P^*=1;
\label{equ:Pstar}
\end{equation}
hence,

\begin{equation}
R^*=6.2773771522\times10^{-5}
\label{equ:Rstar}
\end{equation}
the asterisk meaning "at the equilibrium". 

\par In the orbital model we use, $I_6$ is constant at $0.011204858615$ rad, which is exactly the value of $K^*$ we get. Such an accuracy of 11 digits is given to indicate the numerical equality of the two values, but it has no real physical meaning, so Titan's mean obliquity (measured with respect to its orbital inclination) should be nearly zero. This is confirmed by this formula, given by Henrard and Schwanen (\cite{Henrard04}):

\begin{equation}
K^* \approx \frac{\delta_1+\delta_2}{\delta_1+\delta_2-\frac{\dot{\ascnode}}{n}} I,
\label{equ:hs}
\end{equation}
which becomes $K^* \approx I$ when the mean value of the precession rate of the line of nodes is zero.

\subsection{The fundamental frequencies of the free librations}

\par Since the equilibrium has been found, the Hamiltonian is centered in order to study the behavior of the system near the equilibrium. We introduce a new set of canonical variables: 

\begin{center}
$\begin{array}{lll}
\xi_\sigma=\sigma & \hspace{2.9cm} & \eta_\sigma=P-P^* \\
\xi_\rho=\rho & \hspace{2.9cm} & \eta_\rho=R-R^* \\
\xi_q & \hspace{2.9cm} & \eta_q \\
\end{array}$
\end{center}
As a translation, this transformation is canonical. In these variables, the main part of the Hamiltonian of the problem is quadratic. Its quadratic part is named $\mathcal{N}$ and we have

\begin{equation}
\begin{split}
\frac{\mathcal{N}}{n(1+\delta_s)}=\gamma_{\sigma\sigma}\xi_\sigma^2+2\gamma_{\sigma\rho}\xi_\sigma\xi_\rho+\gamma_{\rho\rho}\xi_\rho^2+\gamma_{qq}\xi_q^2 \\
+\mu_{\sigma\sigma}\eta_\sigma^2+2\mu_{\sigma\rho}\eta_\sigma\eta_\rho+\mu_{\rho\rho}\eta_\rho^2+\mu_{qq}\eta_q^2.
\end{split}
\label{equ:N}
\end{equation}

\par The analytical expressions of the coefficients $\mu_{xx}$ and $\gamma_{xx}$ are recalled in App.\ref{sec:gamu}, and their numerical values are gathered in Table \ref{tab:muga}. The reader should be aware that these coefficients are similar to those in Henrard and Schwanen (\cite{Henrard04}), but different from the ones used by Henrard for Io (\cite{Henrard05i}) and Europa (\cite{Henrard05c}) where other variables are used.

\begin{table}
\centering
\caption{Numerical values of the coefficients $\mu_{xx}$ and $\gamma_{xx}$}
\begin{tabular}{l|l}
Parameter & Numerical Value \\
\hline
$\gamma_{\sigma\sigma}$ & $2.1717941364\times10^{-4}$ \\
$\gamma_{\sigma\rho}$ & $1.3632529077\times10^{-8}$ \\
$\gamma_{\rho\rho}$ & $1.6372888272\times10^{-8}$ \\
$\gamma_{qq}$ & $3.4788181236\times10^{-4}$ \\
$\mu_{\sigma\sigma}$ & $5.0000000409\times10^{-1}$ \\
$\mu_{\sigma\rho}$ & $-6.5206613577\times10^{-5}$ \\
$\mu_{\rho\rho}$ & $1.0387557096$ \\
$\mu_{qq}$ & $1.4564940392\times10^{-5}$ \\
\hline
\end{tabular}
\label{tab:muga}
\end{table}

\par We now introduce the following new set of canonical variables:

\begin{center}
$\begin{array}{lll}
\xi_{\sigma}=x_1-\beta x_2 & \hspace{1.5cm} & \eta_{\sigma}=(1-\alpha\beta)y_1-\alpha y_2 \\
\xi_{\rho}=\alpha x_1+(1-\alpha\beta)x_2 & \hspace{1.5cm} & \eta_{\rho}=\beta y_1+y_2 \\
\xi_q=x_3 & \hspace{1.5cm} & \eta_q=y_3 \\
\end{array}$
\end{center}
with $\alpha$ and $\beta$ conveniently chosen so as to untangle the variables $\xi$ and $\eta$. It can be easily checked that this transformation is canonical, because it preserves the differential form; i.e.

\begin{equation}
d\xi_{\sigma}.\eta_{\sigma}+d\xi_{\rho}.\eta_{\rho}+d\xi_q.\eta_q=dx_1.y_1+dx_2.y_2+dx_3.y_3.
\label{equ:canonical}
\end{equation}
With these new variables, the Hamiltonian (\ref{equ:N}) can be written as

\begin{equation}
\frac{\mathcal{N}}{n(1+\delta_s)}=\zeta_1x_1^2+\zeta_2x_2^2+\zeta_3x_3^2+\psi_1y_1^2+\psi_2y_2^2+\psi_3y_3^2
\label{equ:nuntang}
\end{equation}
with

\begin{equation}
\zeta_1=\gamma_{\sigma\sigma}+2\gamma_{\sigma\rho}\alpha+\gamma_{\rho\rho}\alpha^2
\label{equ:zeta1}
\end{equation}

\begin{equation}
\zeta_2=\gamma_{\sigma\sigma}\beta^2-2\beta(1-\alpha\beta)\gamma_{\sigma\rho}+\gamma_{\rho\rho}(1-\alpha\beta)^2
\label{equ:zeta2}
\end{equation}

\begin{equation}
\psi_1=\mu_{\sigma\sigma}(1-\alpha\beta)^2+2\beta(1-\alpha\beta)\mu_{\sigma\rho}+\beta^2\mu_{\rho\rho}
\label{equ:psi1}
\end{equation}

\begin{equation}
\psi_2=\alpha^2\mu_{\sigma\sigma}-2\alpha\mu_{\sigma\rho}+\mu_{\rho\rho}
\label{equ:psi2}
\end{equation}

\begin{equation}
\zeta_3=\gamma_{qq}
\label{equ:zeta3}
\end{equation}

\begin{equation}
\psi_3=\mu_{qq}.
\label{equ:psi3}
\end{equation}

\par The numerical values of these coefficients are gathered Table \ref{tab:decoupl}.

\begin{table}
\centering
\caption{Numerical values of the coefficients of the Hamiltonian $\mathcal{N}$ after the variables have been untangled.}
\begin{tabular}{l|l}
Parameter & Numerical Value \\
\hline
$\zeta_1$ & $2.1717941364\times10^{-4}$ \\
$\zeta_2$ & $1.6372032547\times10^{-8}$ \\
$\zeta_3$ & $3.4788181236\times10^{-4}$ \\
$\psi_1$ & $0.5000000000$ \\
$\psi_2$ & $1.0387557096$ \\
$\psi_3$ & $1.4564940392\times10^{-5}$ \\
$\alpha$ & $-6.1404734778\times10^{-9}$ \\
$\beta$ & $6.2770815833\times10^{-5}$ \\
\hline
\end{tabular}
\label{tab:decoupl}
\end{table}

\par We can now introduce the last following set of polar canonical coordinates:

\begin{center}
$\begin{array}{lll}
x_1=\sqrt{2UU^*}\sin u & \hspace{2.5cm} & y_1=\sqrt{\frac{2U}{U^*}}\cos u \\
x_2=\sqrt{2VV^*}\sin v & \hspace{2.5cm} & y_2=\sqrt{\frac{2V}{V^*}}\cos v \\
x_3=\sqrt{2WW^*}\sin w & \hspace{2.5cm} & y_3=\sqrt{\frac{2W}{W^*}}\cos w \\
\end{array}$
\end{center}
with

\begin{equation}
U^*=\sqrt{\frac{\psi_1}{\zeta_1}}
\label{equ:Ustar}
\end{equation}

\begin{equation}
V^*=\sqrt{\frac{\psi_2}{\zeta_2}}
\label{equ:Vstar}
\end{equation}

\begin{equation}
W^*=\sqrt{\frac{\psi_3}{\zeta_3}}.
\label{equ:Wstar}
\end{equation}

\par The purpose of this last canonical transformation is to show the free librations around the exact Cassini state. The arguments of these free librations are $u$, $v$, and $w$, and the amplitudes associated are proportional to $\sqrt{U}$, $\sqrt{V}$, and $\sqrt{W}$ respectively. We can easily check that this transformation is canonical because we have $du.U+dv.V+dw.W=dx_1.y_1+dx_2.y_2+dx_3.y_3$.  We can now write

\begin{equation}
\mathcal{N}=\omega_uU+\omega_vV+\omega_wW
\label{equ:Nafk}
\end{equation}
with

\begin{equation}
\omega_u=2n_6\sqrt{\psi_1\zeta_1}(1+\delta_s)
\label{equ:omegu}
\end{equation}

\begin{equation}
\omega_v=2n_6\sqrt{\psi_2\zeta_2}(1+\delta_s)
\label{equ:omegv}
\end{equation}

\begin{equation}
\omega_w=2n_6\sqrt{\psi_3\zeta_3}(1+\delta_s)
\label{equ:omegw}
\end{equation}
\par The numerical results are gathered in Table \ref{tab:frek}, so the periods associated to the 3 free librations around the equilibrium state are respectively $2.09$, $167.37$, and $306.62$ years. Table \ref{tab:comprambh} gives an application of the formulae given in this paper to the Galilean satellites of Jupiter Io and Europa, and we make a comparison with the analytical results of Henrard (\cite{Henrard05i} and \cite{Henrard052}) and the numerical results of Rambaux and Henrard (\cite{Rambaux05}) obtained with the SONYR model (Rambaux \& Bois \cite{Rambaux04}), which is a relativistic N-body model. The small differences between our results and Henrard's analytical results come from Henrard neglecting $a_3$, $b_3$, $b_4$, and $b_5$ for Io, and $b_4$ and $b_5$ for Europa.

\begin{table}[ht]
\centering
\caption{The free librations around the equilibrium state.}
\begin{tabular}{l|cc}
Proper Modes & $\omega$ ($rad.y^{-1}$) & $T$ (period in years) \\
\hline
u & $2.9998383244$ & $2.0945079794$ \\
v & $3.7541492157\times10^{-2}$ & $167.36642435$ \\
w & $2.0491499350\times10^{-2}$ & $306.62399075$ \\
\end{tabular}
\label{tab:frek}
\end{table}

\begin{table}[ht]
\centering
\caption{Comparison of the periods of the free librations of Io and Europa given by different models.}
\begin{tabular}{l|ccc}
Proper Modes & Henrard & SONYR & this paper \\
\hline
Io & & & \\
u & $13.25$ days & $13.18$ days & $13.31$ days \\
v & $159.39$ days & $157.66$ days & $160.20$ days \\
w & $229.85$ days & & $228.53$ days \\
\hline
Europa & & & \\
u & $52.70$ days & $55.39$ days & $52.98$ days \\
v & $3.60$ years & $4.01$ years & $3.65$ years \\
w & $4.84$ years & & $4.86$ years \\
\hline
\end{tabular}
\begin{note}
The results labelled "Henrard" come from (Henrard \cite{Henrard05i}) for Io and (Henrard \cite{Henrard052}) for Europa, while the results labelled "SONYR" come from (Rambaux \& Henrard \cite{Rambaux05}).
\end{note}
\label{tab:comprambh}
\end{table}

\section{Numerical study}

\par To check the reliability of our previous results and to go further in the study of Titan's forced rotation, we used the numerical tool. This allowed us first to obtain a solution for the rotation of Titan and then to describe it by frequency analysis  and to split the free from the forced solutions.

\subsection{Numerical integration}

\par We integrated the 6 equations coming from the Hamiltonian (\ref{equ:hamilton}) over 9000 years, i.e. between -4500 and 4500 years, the time origin being J1980. In these equations, $x$ and $y$ come from TASS1.6 ephemerides. We recall that $x$, $y$, and $z$ are the coordinates of the barycenter of mass of Saturn in the frame $(\vec{f_1},\vec{f_2},\vec{f_3})$ rigidly linked to Titan (see Sect. 2.3).

\par We used the Adams-Bashforth-Moulton $10^{th}$-order predictor-corrector integrator, with a constant timestep $h=1.6\times10^{-4}$ year, i.e. $5.844\times10^{-2}$ day. We considered that the shortest significant fundamental period of the system is given by $3\lambda_6$, i.e. $\approx 5.315$ days $\approx 90\times h$.

\begin{table}
\centering
\caption{Initial conditions chosen for the numerical integration, at t=-4500 years.}
\begin{tabular}{l|c}
Variable & Expression \\
\hline
$p_0$ & $\lambda_{60}-\pi$ \\
$r_0$ & $-\ascnode_{60}$ \\
$\xi_0$ & $10^{-4}$ \\
$\eta_0$ & $10^{-4}$ \\
$P_0$ & $1-\frac{\dot{\ascnode}_{60}}{n_6}(1-\cos K_0)+10^{-4}$ \\
$R_0$ & $1.0001\times P_0(1-\cos K_0)$ \\
\hline
\end{tabular}
\label{tab:condini1}
\begin{note}
These conditions have been arbitrarily chosen near the Cassini state, with $\lambda_{60}$, $\ascnode_{60}$, and $\dot{\ascnode}_{60}$ respectively the values of Titan's mean longitude, argument of the ascending node, and its instantaneous angular rate, at t=-4500 years, given by TASS1.6. $K_0$ is the initial value of the obliquity $K$ on the same date.
\end{note}
\end{table}

\par Table \ref{tab:condini1} gathers the initial conditions we used. These conditions were arbitrarily chosen near the Cassini state. It implies that, by choosing these initial conditions, we supposed that Titan is at the Cassini state. In these initial conditions, the initial value of $K$ $K_0$ is defined as 

\begin{equation}
K_0=\frac{\delta_1+\delta_2}{\delta_1+\delta_2-\frac{\dot{\ascnode}_{60}}{n_6}}.
\label{equ:kzero}
\end{equation}
This equation is very similar to (\ref{equ:hs}). We used it to be sure that the system is near the equilibrium. We did not want to start at the exact equilibrium but very close in order to be able to detect the 3 free librations that we studied in the previous section. However, we should keep in mind that the frequencies computed in Table \ref{tab:frek} are in fact limits of the frequencies of the free librations when their amplitudes tend to zero. Thus, too high amplitudes of free librations would alter the frequencies too much. In that way, their comparison with the expected fundamental frequencies of the free librations might be difficult, so their identification as these fundamental frequencies could become doubtful.

\begin{figure*}[ht]
\centering
\begin{tabular}{cc}
\includegraphics[width=8.7cm]{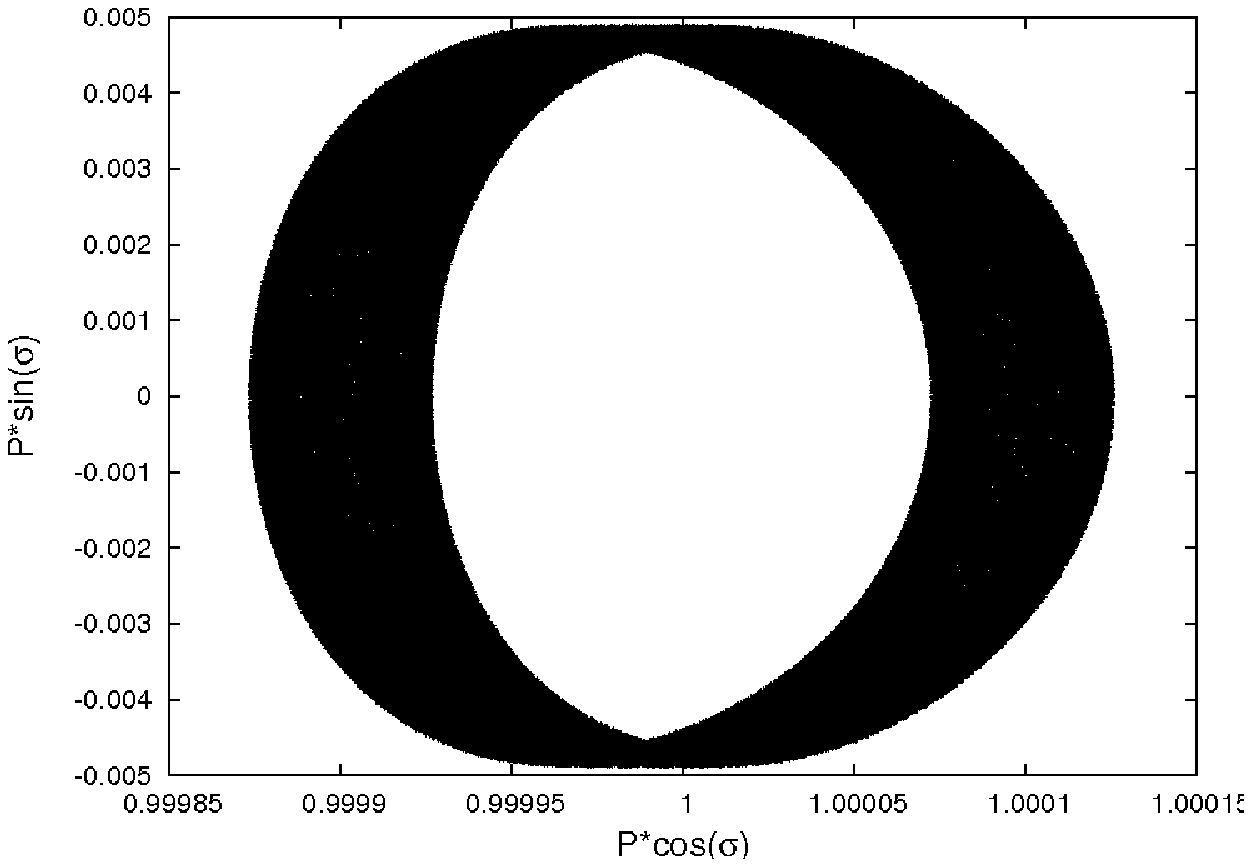} & \includegraphics[width=8.7cm]{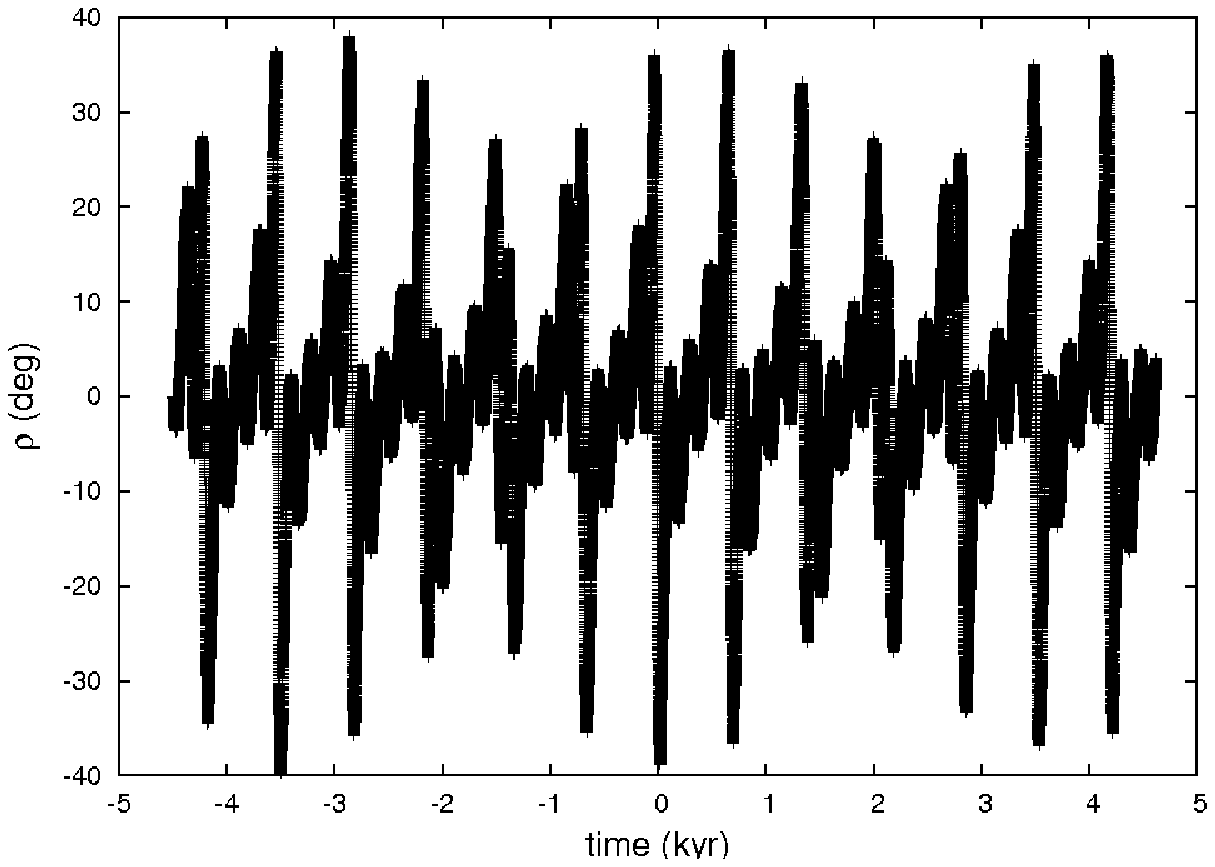} \\
(a) & (b) \\
\includegraphics[width=8.7cm]{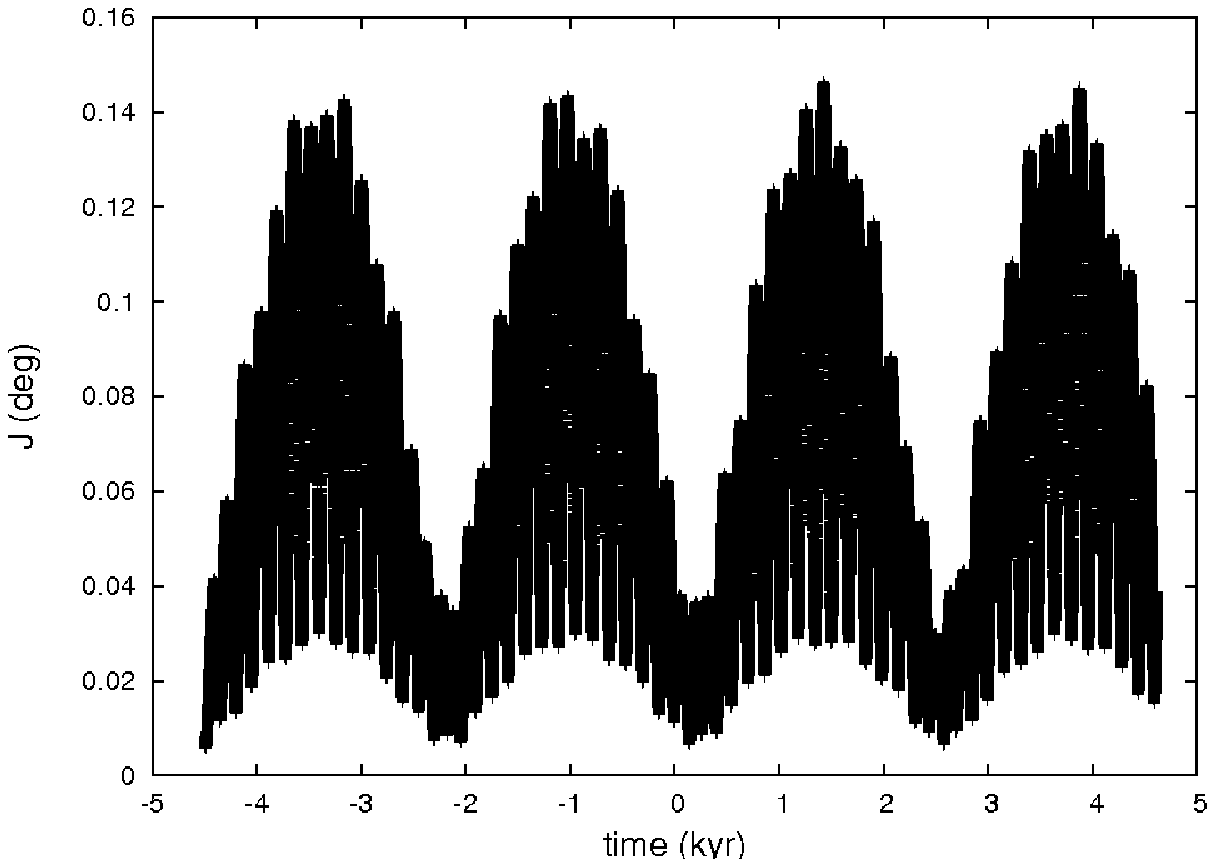} & \includegraphics[width=8.7cm]{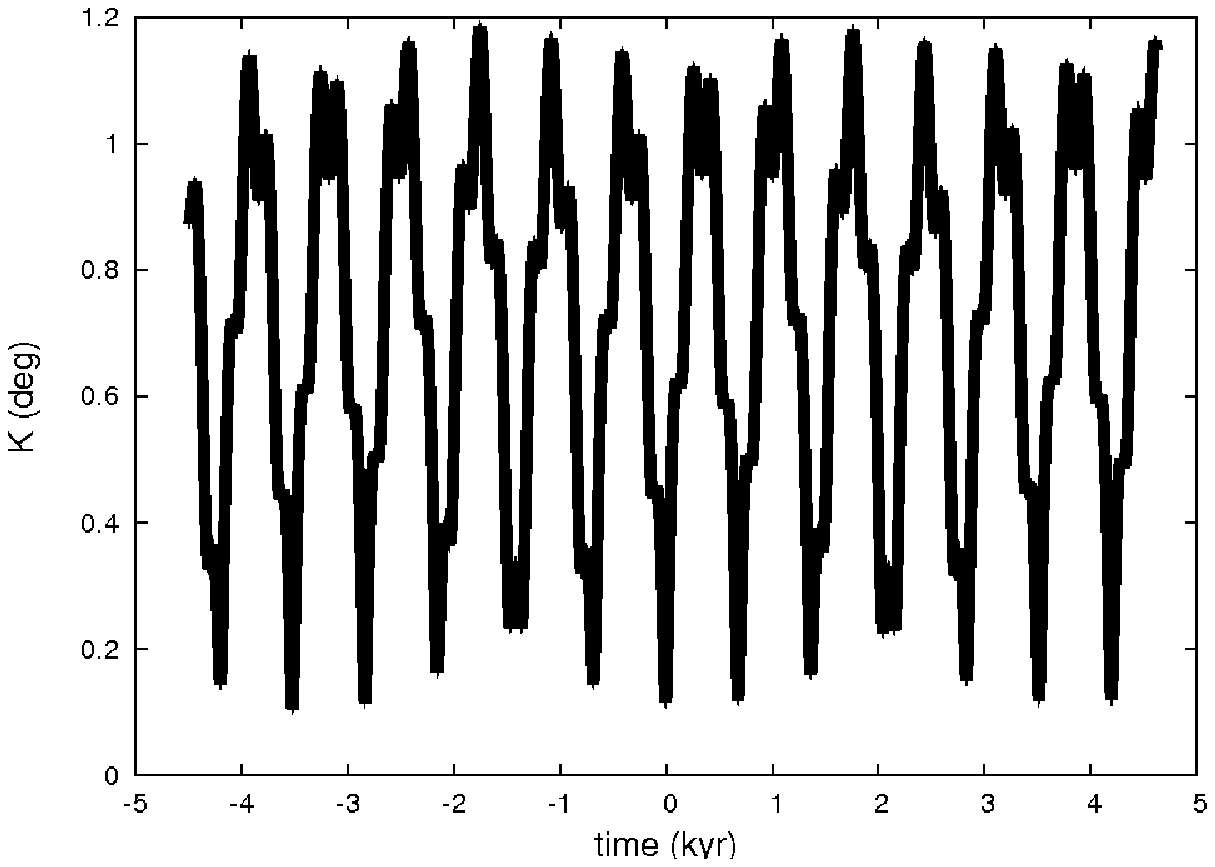} \\
(c) & (d) \\
\end{tabular}
\caption{Numerical simulation of Titan's obliquity over 9000 years, the time origin being J1980= 2444240 JD. Here the behavior of the variables $P$, $\sigma$, $\rho$, $J$ and $K$ is being displayed. Explanations are in the text.}
\label{fig:tithr}
\end{figure*}

\par Figure \ref{fig:tithr} gives plots of some significant data resulting from the numerical integration. Figure \ref{fig:tithr}a shows the behavior of the variables $P=\frac{G}{nC}$ (modulus of the plotted value) and $\sigma$, and Figure \ref{fig:tithr}b shows the behavior of $\rho$, i.e. the difference between the two nodes. We can see that this angle is oscillating around $0$, as predicted by the theory. We can also visually detect a period of about $700$ years, and will see later that it is a forced component due to the behavior of Titan's orbital ascending node. Figure \ref{fig:tithr}c shows the behavior of the wobble $J$, we obtained it from the variables $\xi_q$ and $\eta_q$. Finally, Figure \ref{fig:tithr}d shows the ``obliquity'' $K$. In this last panel, we can see the same 700-year-periodic contribution detected in $\rho$. Unfortunately, looking at these plots does not give information on the free and the forced components of the solutions. That is why we used the frequency analysis technique.

\subsection{Analysis of the solutions}

\par We use the frequency analysis to describe the solutions given by the numerical integration, i.e. to give a quasi-periodic representation of these solutions. Such a technique has already been used often to describe the orbital motion of planets (see Laskar \cite{Laskar88}) or natural satellites (cf. for instance Vienne \& Duriez \cite{Vienne95} or very recently Lainey et al. \cite{Lainey06}).

\par One of the main difficulties with this kind of problem is that we have two timescales for the periods of the terms that appear in the synthetic representations: Titan's orbital period is roughly $16$ days, while the period of its pericenter is about $700$ years. In order to correctly detect the long-period terms, the total time-interval used to analyze the solution should be about as long as the longest period expected, here $\approx 3200$ years, and the timestep shorter than half the shortest period expected (about 5 days). Thus, data over 3200 years should be represented with a timestep of 2.5 days, but this would require about 500000 points. This would take a very long computation time, but fortunately some alternative techniques exist to solve this problem.

\par The most common technique is the use of a digital filter that splits the short-period terms from the long-period ones (see Carpino et al. \cite{Carpino87}). However, this technique might alter the signal. Another technique has been used here, which consists in using two samples of data with very close timesteps, as explained in Laskar (\cite{Laskar04}). More precisely, for each variable, we extracted two samples of 65536 data from the results of the numerical integration, one point every 848 for the first sample and one point every 864 for the other one. As a result, the first sample represents the solutions for $8891.7888$ years with a timestep $h_1=49.55712$ days, and the second represents the solutions for $9059.5584$ years with a timestep $h_2=50.49216$ days.

\par These two timesteps are far too large to detect contributions with a period of about $16$ days. In fact, the short-period terms are detected, but with a wrong frequency. When a frequency $\nu$ is too high, it is detected as 

\begin{equation}
\nu_1=\nu+\frac{k_1}{h_1}
\label{equ:nu1}
\end{equation}
in analyzing the first sample, and as 

\begin{equation}
\nu_2=\nu+\frac{k_2}{h_2}
\label{equ:nu2}
\end{equation}
in analyzing the second one, where $k_1$ and $k_2$ are (a priori unknown) integers. We have

\begin{equation}
(\nu_2-\nu_1)h_2=k_2-k_1\frac{h_2}{h_1}
\label{equ:diff1}
\end{equation}
and

\begin{equation}
\nu_2h_2-\nu_1h_1=\nu(h_2-h_1)+k_2-k_1.
\label{equ:diff2}
\end{equation}
If we now define $[x]$ as the closest integer to the real $x$ (i.e. $|[x]-x|<\frac{1}{2}$), we have 

\begin{equation}
[\nu_2h_2-\nu_1h_1]=k_2-k_1
\label{equ:diff3}
\end{equation}
and finally

\begin{equation}
k_1=\frac{h_2}{h_1-h_2}((\nu_2-\nu_1)h_2-[\nu_2h_2-\nu_1h_1]),
\label{equ:k1}
\end{equation}
where Eq.(\ref{equ:diff3}) requires that $h_1$ and $h_2$ are close enough, i.e. $|\nu(h_2-h_1)|<\frac{1}{2}$. In our case, the highest frequency that we can detect with this method is $\frac{1}{2(h_2-h_1)} \approx 0.54 d^{-1}$, so we can detect every term with a period longer than $11.75$ days, while analyzing only one sample would give periods longer than about 100 days (i.e. 2 timesteps). Such accuracy is enough to detect Titan's orbital period.

\par Table \ref{tab:Ptits1} is an example of the decomposition of a variable (here $P=\frac{G}{nC}$, with $G$ the norm of the angular momentum) with the two timesteps. The algorithm used for determining each frequency is taken from (Laskar et al. \cite{Laskar92}) and has been iteratively applied to refine each frequency, as described in (Champenois \cite{Champenois98}). In this table, term 1 is a constant part, while the second one is clearly the free libration associated to the proper mode $u$. The slight difference between the obtained and the expected periods should be partly due to the associated amplitude not being null, and partly to the approximations used in our analytical model (i.e. no eccentricity and a constant inclination). However, we can see that the two determinations give very different results for term 3, so we can infer that it is in fact a short-period term.

\begin{table*}[ht]
\centering
\caption{Decomposition of the solution for $P$ with the two timesteps, i.e. $h_1=49.55712$ days (left) and $h_2=50.49216$ days (right).}
\begin{tabular}{r|ccc|ccc}
N\textdegree & Amp. & Phase (\textdegree) & T (y) & Amp. & Phase (\textdegree) & T (y) \\
\hline
1 & $1.000000002$ & $-4.57\times10^{-8}$ & $3.50\times10^{13}$ & $1.000000002$ & $8.95\times10^{-8}$ & $-1.82\times10^{13}$ \\
2 & $0.000099514$ & $0.76$ & $2.09773$ & $0.000099514$ & $0.77$ & $2.09773$ \\
3 & $0.000025104$ & $144.00$ & $1.25952$ & $0.000025104$ & $144.00$ & $0.83097$ \\
\hline
\end{tabular}
\begin{note}
The origin of phases is here the origin of the frequency analysis, i.e. 4499.99344 years before J1980. The series are given in cosine.
\end{note}
\label{tab:Ptits1}
\end{table*}

\par Applying (\ref{equ:k1}) we find $T=15.6612$ days. That is very near to Titan's orbital period, so this term should be an integer combination of Titan's orbital period and other contribution(s).

\par It is now interesting to identify the periodic terms contained in the solutions associated to the considered variables. Table \ref{tab:propmodes} gives the proper modes that are expected. They should appear in the quasiperiodic decompositions of the solutions as parts of integer combinations, so integer combinations of the frequencies of the proper modes are performed to identify each term of the decompositions. We do not use the phases because they are uncertain in the Titan ephemerides given by Vienne \& Duriez (\cite{Vienne95}). The reason is that the given phases are in fact integer combinations of the phases coming from the identified proper modes and very-long-period arguments due to the solar perturbation that are assumed to be constant on an ephemerides-timescale.

\begin{table}[ht]
\centering
\caption{Proper modes of the system.}
\begin{tabular}{llll}
\hline
Proper & Frequency & Period & Cause \\
Mode & $(rad.y^{-1})$ & & \\
\hline
$\lambda_5$ & $508.00932017$ & $4.52$ days & Rhea \\
$\lambda_6$ & $143.92404729$ & $15.95$ days & Titan \\
$\lambda_8$ & $28.92852233$ & $79.33$ days & Iapetus \\
$\phi_5$ & $0.17554922$ & $35.79$ years & $e_5$ \\
$\Phi_5$ & $-0.17546762$ & $35.81$ years & $\gamma_5$ \\
$\phi_6$ & $0.00893386$ & $703.30$ years & $e_6$ \\
$\Phi_6$ & $-0.00893124$ & $703.51$ years & $\gamma_6$ \\
$\phi_8$ & $0.00197469$ & $3181.86$ years & $e_8$ \\
$\Phi_8$ & $-0.00192554$ & $3263.07$ years & $\gamma_8$ \\
$\lambda_9$ & $0.21329912$ & $29.46$ years & Sun \\
\hline
$\phi_u$ & $2.995$ & $2.09773$ years & $\sqrt{U}$ \\
$\phi_v$ & $0.0375$ & $167.4883$ years & $\sqrt{V}$ \\
$\phi_w$ & $0.0205$ & $306.3360$ years & $\sqrt{W}$ \\
\hline
\end{tabular}
\begin{note}
The modes $\lambda_5$ to $\lambda_9$ (first part of the Table) come from Vienne \& Duriez (\cite{Vienne95}), while the second part contains the free librations around the Cassini state. These terms have been evaluated from the solutions given by our numerical integration. The fourth column gives the orbital parameter to which the proper mode is linked, $e_i$ being the eccentricity of the satellite $i$, and $\gamma_i$ the sine of its semiinclination. The subscripts $i$ are $5$ for Rhea, $6$ for Titan, and $8$ for Iapetus.
\end{note}
\label{tab:propmodes}
\end{table}

\begin{table*}[ht]
\centering
\caption{Quasiperiodic decomposition of the variable $P$.}
\begin{tabular}{r|ccccc}
N\textdegree & Amp. & Phase (\textdegree) & T (y) & Ident. & Cause \\
\hline
1 & $1.000000002$ & $4.89\times10^{-10}$ & $-1.82\times10^{13}$ & constant & \\
2 & $0.000099514$ & $63.00$ & $2.09773$ & $\phi_u$ & $\sqrt{U}$ \\
3 & $0.000025104$ & $35.44$ & $15.6612$ days & unknown & \\
\hline
\end{tabular}
\begin{note}
The series are in cosine. The fourth column gives the orbital parameters associated to each identified term.
\end{note}
\label{tab:P}
\end{table*}

\begin{table*}[ht]
\centering
\caption{Quasiperiodic decomposition of the variable $R$. The series are in cosine.}
\begin{tabular}{r|ccccc}
N\textdegree & Amp. $\times10^5$ & Phase (\textdegree) & T (y) & Ident. & Cause \\
\hline
1 & $9.17912502$ & $2.07\times10^{-6}$ & $6.12\times10^{10}$ & constant & \\
2 & $8.22242693$ & $-170.93$ & $703.50790$ & $-\Phi_6$ & $\gamma_6$ \\
3 & $2.28952572$ & $174.72$ & $167.48834$ & $\phi_v$ & $\sqrt{V}$ \\
4 & $1.49857669$ & $-143.60$ & $219.82166$ & $\Phi_6+\phi_v$ & $\sqrt{V}\gamma_6$ \\
5 & $0.30469561$ & $-107.54$ & $3252.81$ & $\Phi_8$ & $\gamma_8$ \\
6 & $0.22920477$ & $-61.85$ & $899.49195$ & $\Phi_8-\Phi_6$ & $\gamma_6\gamma_8$ \\
7 & $0.05732146$ & $-80.14$ & $176.5406$ & $\Phi_8+\phi_v$ & $\sqrt{V}\gamma_8$ \\
8 & $0.01907471$ & $163.85$ & $14.72857$ & $2\lambda_9$ & Sun \\
\hline
\end{tabular}
\label{tab:R}
\end{table*}

\begin{table*}[ht]
\centering
\caption{Quasiperiodic decomposition of the complex variable $\eta_q+\sqrt{-1}\xi_q$.}
\begin{tabular}{r|ccccc}
N\textdegree & Amp. $\times10^4$ & Phase (\textdegree) & T (y) & Ident. & Cause \\
\hline
1 & $9.12391728$ & $-51.69$ & $306.33602$ & $\phi_w$ & $\sqrt{W}$ \\
2 & $6.01688587$ & $51.69$ & $-306.33605$ & $-\phi_w$ & $\sqrt{W}$ \\
3 & $5.73033451$ & $158.48$ & $351.70284$ & $\phi_6-\Phi_6$ & $e_6\gamma_6$ \\
4 & $3.83212940$ & $-158.48$ & $-351.70284$ & $\Phi_6-\phi_6$ & $e_6\gamma_6$ \\
5 & $0.63642954$ & $-35.86$ & $135.27368$ & $\phi_v-\Phi_6$ & $\sqrt{V}\gamma_6$ \\
6 & $0.38395548$ & $35.86$ & $-135.27368$ & $\Phi_6-\phi_v$ & $\sqrt{V}\gamma_6$  \\
\hline
\end{tabular}
\label{tab:etaxi}
\end{table*}

\begin{table*}[ht]
\centering
\caption{Quasiperiodic decomposition of the variable $\sigma$. The series are in sine.}
\begin{tabular}{r|ccccc}
N\textdegree & Amp. $\times10^3$ & Phase (\textdegree) & T (y) & Ident. & Cause \\
\hline
1 & $4.78176461$ & $63.00$ & $2.09773$ & $\phi_u$ & $\sqrt{U}$ \\
2 & $0.02510524$ & $35.43$ &  $15.6612$ days & unknown & \\
3 & $0.01147635$ & $-5.61$ & $167.47831$ & $\phi_v$ & $\sqrt{V}$ \\
4 & $0.01014094$ & $8.48$ & $703.51797$ & $-\Phi_6$ & $\gamma_6$ \\
5 & $0.00961527$ & $-74.09$ & $55.1128$ days & unknown &  \\
6 & $0.00896290$ & $-56.74$ & $2.11032$ & $\phi_u+2\Phi_6$ & $\sqrt{U}\gamma_6^2$ \\
7 & $0.00891238$ & $2.47$ & $2.08529$ & $\phi_u-2\Phi_6$ & $\sqrt{U}\gamma_6^2$  \\
8 & $0.00744233$ & $137.13$ & $15.6602$ days & unknown & \\
9 & $0.00566630$ & $165.69$ & $219.75766$ & $\phi_v+\Phi_6$ & $\sqrt{V}\gamma_6$ \\
\hline
\end{tabular}
\label{tab:sig}
\end{table*}

\begin{table*}[ht]
\centering
\caption{Quasiperiodic decomposition of the variable $\rho$. The series are in sine.}
\begin{tabular}{r|ccccc}
N\textdegree & Amp. & Phase (\textdegree) & T (y) & Ident. & Cause \\
\hline
1 & $0.18089837$ & $175.64$ & $167.49723$ & $\phi_v$ & $\sqrt{V}$ \\
2 & $0.15667339$ & $-170.90$ & $703.52446$ & $-\Phi_6$ & $\gamma_6$ \\
3 & $0.11829380$ & $-175.17$ & $135.28724$ & $\phi_v-\Phi_6$ & $\sqrt{V}\gamma_6$ \\
4 & $0.09023900$ & $-161.91$ & $351.75789$ & $2\Phi_6$ & $\gamma_6^2$ \\
5 & $0.07735641$ & $-166.02$ & $113.46712$ & $\phi_v-2\Phi_6$ & $\sqrt{V}\gamma_6^2$ \\
6 & $0.05226443$ & $-152.60$ & $234.50407$ & $-3\Phi_6$ & $\gamma_6^3$ \\
7 & $0.05058400$ & $-156.89$ & $97.70793$ & $\phi_v-3\Phi_6$ & $\sqrt{V}\gamma_6^3$ \\
8 & $0.03311443$ & $-147.68$ & $85.79329$ & $\phi_v-4\Phi_6$ & $\sqrt{V}\gamma_6^4$ \\
9 & $0.03060799$ & $-143.30$ & $175.88361$ & $-4\Phi_6$ & $\gamma_6^4$ \\
10 & $0.02165111$ & $-138.40$ & $76.46679$ & $\phi_v-5\Phi_6$ & $\sqrt{V}\gamma_6^5$ \\
11 & $0.02035272$ & $-173.58$ & $74.84659$ & $2\phi_v-\Phi_6$ & $V\gamma_6$ \\
12 & $0.02027563$ & $-167.25$ & $67.64828$ & $2\phi_v-2\Phi_6$ & $V\gamma_6^2$ \\
13 & $0.01799820$ & $-132.60$ & $140.70722$ & $-5\Phi_6$ & $\gamma_6^5$ \\
14 & $0.01785634$ & $-158.19$ & $61.71406$ & $\phi_v-3\Phi_6$ & $\sqrt{V}\gamma_6^3$ \\
15 & $0.01739585$ & $163.87$ & $14.72858$ & $2\lambda_9$ & Sun \\
16 & $0.01498249$ & $-176.87$ & $83.76245$ & $2\phi_v$ & $V$ \\
17 & $0.01472909$ & $-149.81$ & $56.73966$ & $2\phi_v-4\Phi_6$ & $V\gamma_6^4$ \\
18 & $0.01417957$ & $-128.79$ & $68.97041$ & $\phi_v-6\Phi_6$ & $\sqrt{V}\gamma_6^6$ \\
19 & $0.01152268$ & $-140.62$ & $52.50108$ & $2\phi_v-5\Phi_6$ & $V\gamma_6^5$ \\
20 & $0.01065513$ & $-121.56$ & $117.25730$ & $-6\Phi_6$ & $\gamma_6^6$ \\
21 & $0.00929051$ & $-148.93$ & $62.81242$ & unknown & \\
22 & $0.00899501$ & $154.97$ & $15.04352$ & $2\lambda_9+\Phi_6$ & Sun, $\gamma_6$ \\
23 & $0.00881939$ & $-131.88$ & $48.85511$ & $2\phi_v-6\Phi_6$ & $V\gamma_6^6$ \\
24 & $0.00657942$ & $-122.55$ & $45.68121$ & $2\phi_v-7\Phi_6$ & $V\gamma_6^7$ \\
25 & $0.00635299$ & $-110.18$ & $100.50834$ & $-7\Phi_6$ & $\gamma_6^7$ \\
26 & $0.00609640$ & $-109.01$ & $57.66387$ & unknown & \\
27 & $0.00584381$ & $-87.12$ & $109.64279$ & $\phi_v-2\Phi_6-\Phi_8$ & $\sqrt{V}\gamma_6^2\gamma_8$ \\
28 & $0.00584403$ & $-95.53$ & $129.88963$ & $\phi_v-\Phi_6-\Phi_8$ & $\sqrt{V}\gamma_6\gamma_8$ \\
29 & $0.00531164$ & $-19.68$ & $29.44635$ & $\lambda_9$ & Sun \\
30 & $0.00513296$ & $-81.36$ & $317.28640$ & $-2\Phi_6-\Phi_8$ & $\gamma_6^2\gamma_8$ \\
\hline
\end{tabular}
\label{tab:rho}
\end{table*}

\begin{table*}[ht]
\centering
\caption{Quasiperiodic decomposition of $K$. The series are in cosine.}
\begin{tabular}{r|ccccc}
N\textdegree & Amp. $\times10^2$ (rad)  & Phase (\textdegree) & T (y) & Ident. & Cause \\
\hline
1 & $1.25481164$ & $8.68\times10^{-10}$ & $-2.65\times10^{13}$ & constant & \\
2 & $0.68465799$ & $-170.92$ & $703.51272$ & $-\Phi_6$ & $\gamma_6$ \\
3 & $0.17842225$ & $175.02$ & $167.49146$ & $\phi_v$ & $\sqrt{V}$ \\
4 & $0.10246867$ & $-161.88$ & $351.76856$ & $-2\Phi_6$ & $\gamma_6^2$ \\
5 & $0.07264971$ & $-15.67$ & $219.80041$ & $\phi_v+\Phi_6$ & $\sqrt{V}\gamma_6$ \\
\hline
\end{tabular}
\label{tab:K}
\end{table*}

\par The results are summarized in Tables \ref{tab:P} to \ref{tab:K}, the origin of the phases now being J1980. There, $K$ is given in radians, and the other variables have no unit. The terms with a period $T$ written in years could have been obtained in analyzing only one set of data. In fact, the two sets have been analyzed, and the results are the same for these terms. Except for two of them, all the components have been clearly identified. The terms whose periods are written in days were determined by comparing the results given by the two analysis. These terms are not clearly identified, the reason probably being that they require a high accuracy in their determination. In (\ref{equ:k1}), two quantities are substracted, so a cancellation problem might appear and complicate the determination. Moreover, an integer combination between a short-period term (like Titan's mean longitude $\lambda_6$) and a long-period term gives a short-period term very close to the original short period, so the short-period terms that we detected might in fact be sums of several terms with very close frequencies, making them very difficult to split.

\par These short-period terms seem to have a period very close to Titan's orbital period, except the term 5 in the decomposition of $\sigma$. If we consider that the timesteps $h_1$ and $h_2$ are not close enough and that we have in fact $[\nu(h_2-h_1)]=1$, (\ref{equ:k1}) becomes

\begin{equation}
k_1=1+\frac{h_2}{h_1-h_2}((\nu_2-\nu_1)h_2-[\nu_2h_2-\nu_1h_1]),
\label{equ:k11}
\end{equation}
and we obtain a term whose period is $5.22008$ days. This is quite close to the period associated to $3\lambda_6$.

\par The quasiperiodic decomposition of the solutions allows us to split the forced solution away from the free one. The free solution around the equilibrium can only be known with observations that could give initial conditions for the numerical integration. However, the forced solution only depends on the equilibrium and can be obtained in dropping, in the solutions given in Tables \ref{tab:P} to \ref{tab:K}, the terms depending on the free libration modes $\phi_u$, $\phi_v$ and $\phi_w$. Thus, we can for instance see that angle $J$ is not zero at the equilibrium but has a forced motion. This possibility of a forced wobble has already been pointed out by Bouquillon et al. (\cite{Bouquillon03}) in a general study of the rotation of the synchronous bodies (i.e. that are in a $1:1$ spin-orbit resonance).

\section{Discussion}

\subsection{Comparison between the analytical and the numerical results}

\par Table \ref{tab:comptit} gives a comparison between our analytical and numerical results. We recall that, in the analytical model, the orbit of Titan is circular with a constant inclination, whereas the orbital eccentricity of Titan (i.e. $0.0289$) is taken into account in the numerical model, along with the variation in its inclination. We can see very good matching for the periods of the free librations around the equilibrium. In contrast, we can see a significant difference in the equilibrium obliquity $K^*$. The line $\epsilon$ refers to the equilibrium obliquity with the normal of Titan's orbit as its origin. It is computed by substracting the mean inclination of Titan to $K^*$. The mean inclination of Titan is $1.12049\times10^{-2}$ in the analytical model and $1.18985\times10^{-2}$ in the numerical one. The difference in $K^*$ partly comes from the difference in the mean inclination of Titan, but probably not only from it.

\begin{table*}[ht]
\centering
\caption{Comparison between our analytical and numerical results.}
\begin{tabular}{l|ccc}
\hline
Parameter & Analytical & Numerical & Difference \\
\hline
$K^*$ (rad) & $1.1204859\times10^{-2}$ & $1.25481164\times10^{-2}$ & $12\%$ \\
$\epsilon$ (arcmin) & $0$ & $2.233$ & $(\ldots)$ \\
$T_u$ (y) & $2.094508$ & $2.09773$ & $0.15\%$ \\
$T_v$ (y) & $167.36642$ & $167.49723$ & $0.08\%$ \\
$T_w$ (y) & $306.62399$ & $306.33602$ & $0.09\%$ \\
\hline
\end{tabular}
\label{tab:comptit}
\end{table*}

\subsection{Influence of Titan's inclination and eccentricity}

\par Titan's inclination plays an overwhelming role in its obliquity, as shown in (\ref{equ:hs}). Moreover, the proper modes $\Phi_5$, $\Phi_6$, and $\Phi_8$ given in Table \ref{tab:propmodes} can be linked in its inclination, because they consist of the main (or at least second) part of the solutions for $\zeta=\sin\frac{I}{2}exp(\sqrt{-1}\ascnode)$ for Rhea, Titan, and Iapetus in TASS1.6, and they appear in the solution for $\zeta_6$ (related to Titan's inclination). It is striking, for instance in reading Table \ref{tab:rho}, that the term $\Phi_6$ plays a very important role in the forced and in the free solution. We can even figure the period of $703.51$ years just in looking at Fig.\ref{fig:tithr}b. 

\par In contrast, the proper modes $\phi_5$, $\phi_6$, and $\phi_8$ do not clearly appear, with the exception of $\phi_6$ in $\eta_q+\sqrt{-1}\xi_q$. These modes are related to the eccentricities of Rhea, Titan, and Iapetus, and their values are respectively $10^{-3}$, $0.0289$, and $0.0294$. In fact, $\phi_6$ might play a more important role than suggested by Tables \ref{tab:P} to \ref{tab:K}, because it could be confused with $-\Phi_6$ by the algorithm of frequency analysis. The reason is that these two terms have very close periods, i.e. $703.3$ and $703.51$ years. If we call $\nu_1$ and $\nu_2$ the associated frequencies and $\nu_0$ the Fourier fundamental frequencies (i.e. the frequency associated to a term whose period is the interval of study, 9000 years in our cases, it has no link to the fundamental frequencies of the system), the algorithm of frequency analysis can split $\nu_1$ from $\nu_2$ only if

\begin{equation}
|\nu_1-\nu_2|>2\nu_0.
\label{equ:splitNAFF}
\end{equation}
This implies that the interval of study should be longer than $4.712 \times10^6yr$. Such a timescale is not consistent with the ephemerides and so cannot be considered. It does not mean that the terms identified as $\Phi_6$ are in fact $\phi_6$, because the period found is much closer to $703.51$ years than to $703.30$. It just means that there might be a very small contribution due to $\phi_6$ in the identified term. Moreover, we cannot exclude a role played by the eccentricities in the values of the equilibrium obliquity and of the fundamental frequencies of the free librations.

\subsection{Uncertainty on Titan's gravitational field}

\par Titan's gravitational field is not clearly known. We are confident in its mass thanks to the Pioneer and Voyager fly-bys (see Campbell \& Anderson \cite{Campbell89}), but we are uncertain about $10\%$ of its $J_2$, and we have no value for the ratio $\frac{C}{MR^2}$. We can just hypothesize that it is included between $0.3$ and $0.4$, as it is the case for the Galilean satellites for Jupiter. We arbitrarily chose $\frac{C}{MR^2}=0.31$ and also tried with $\frac{C}{MR^2}=0.35$.

\begin{table}[ht]
\centering
\caption{The free librations around the equilibrium state, with $\frac{C}{MR^2}=0.35$}
\begin{tabular}{l|cc}
Proper Modes & $\omega$ ($rad.y^{-1}$) & $T$ (period in years) \\
\hline
u & $2.822839$ & $2.225839$ \\
v & $3.324655\times10^{-2}$ & $188.987571$ \\
w & $1.814709\times10^{-2}$ & $346.236493$ \\
\end{tabular}
\label{tab:frekdr}
\end{table}

\begin{figure*}[ht]
\centering
\begin{tabular}{cc}
\includegraphics[width=8.7cm]{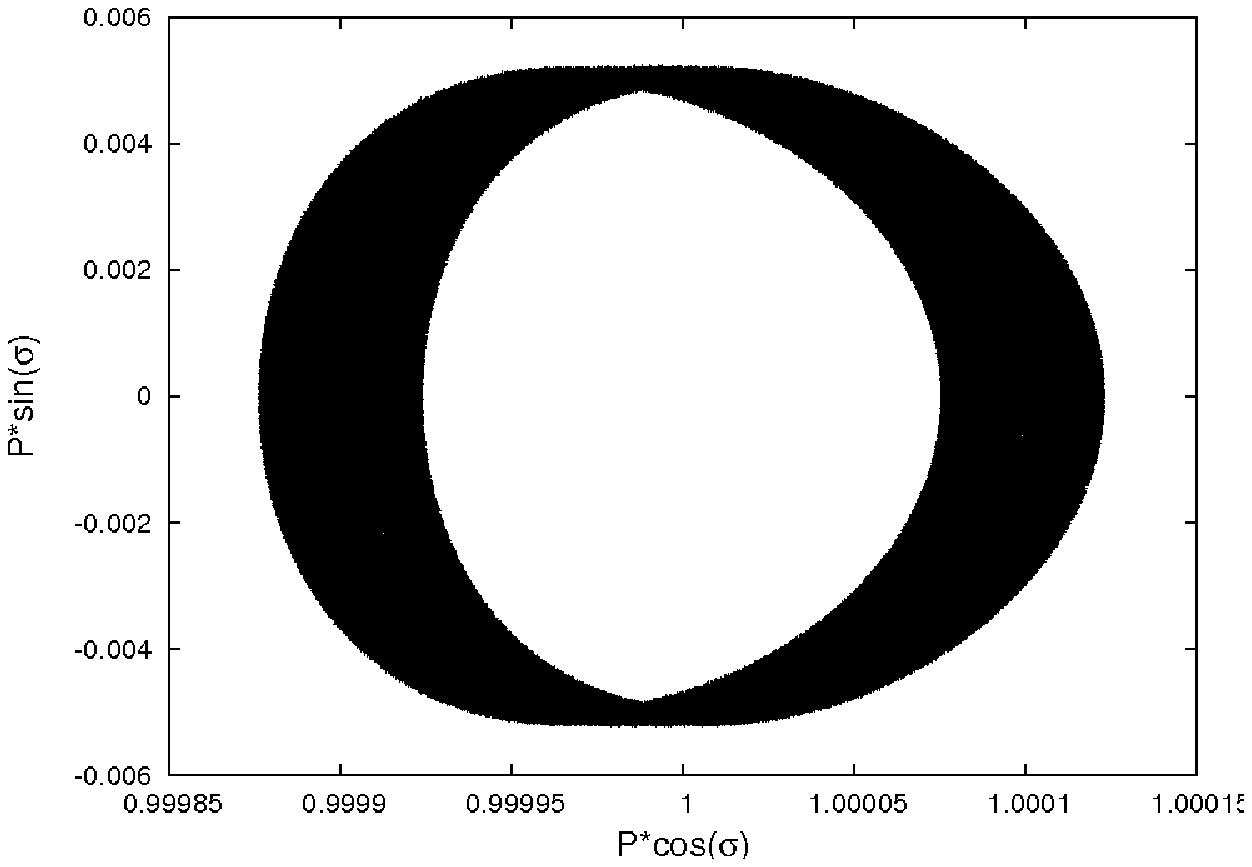} & \includegraphics[width=8.7cm]{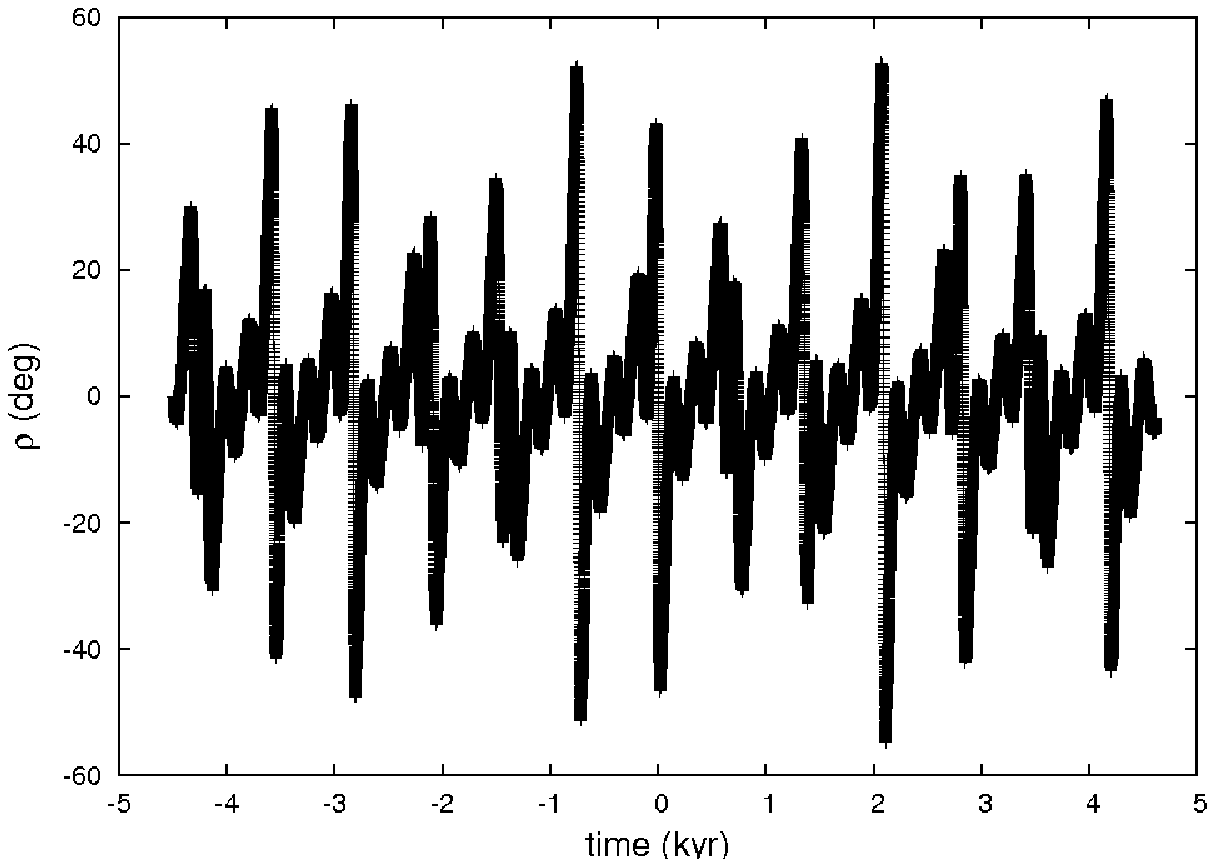} \\
(a) & (b) \\
\includegraphics[width=8.7cm]{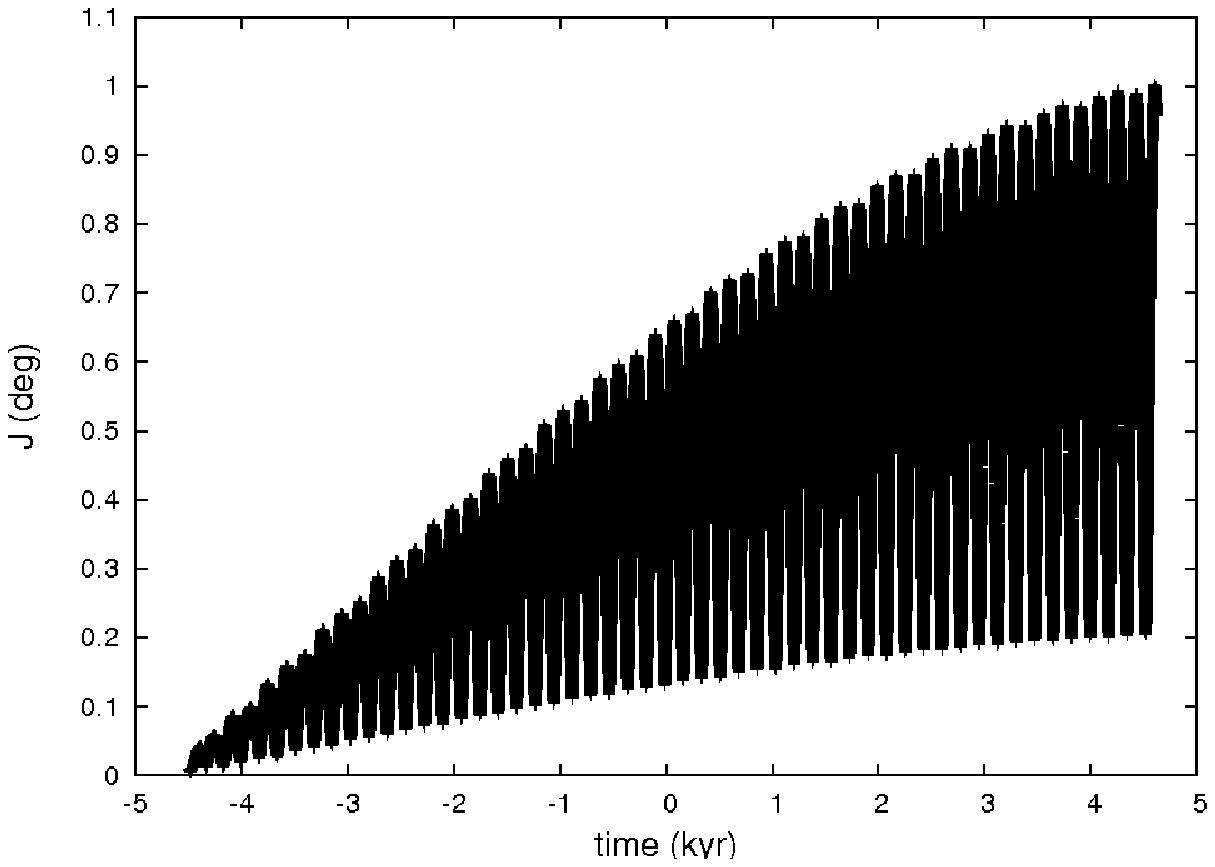} & \includegraphics[width=8.7cm]{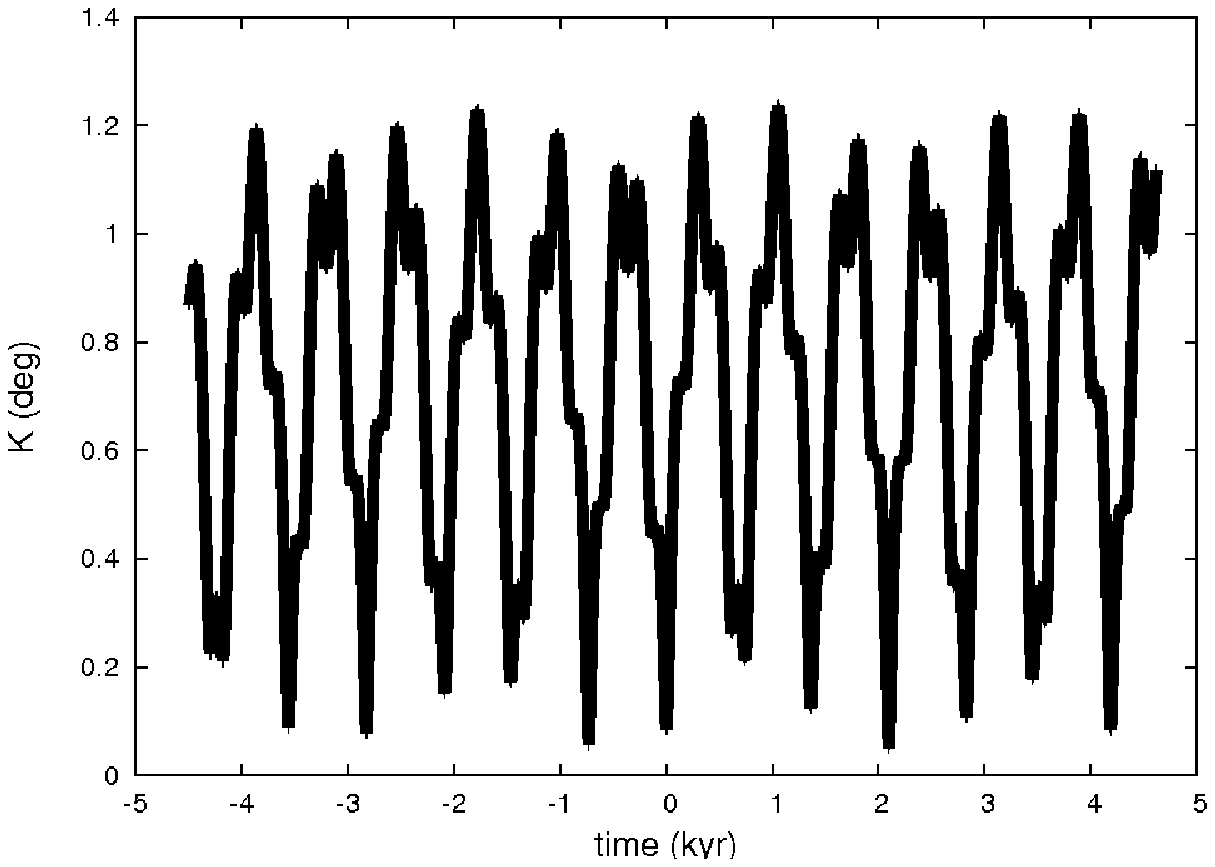} \\
(c) & (d) \\
\end{tabular}
\caption{Numerical simulation of Titan's obliquity over 9000 years, with $\frac{C}{MR^2}=0.35$. The displayed variables are the same as in Fig.\ref{fig:tithr}.}
\label{fig:titdr}
\end{figure*}

\begin{table*}[ht]
\centering
\caption{Comparison between our analytical and numerical results, with $\frac{C}{MR^2}=0.35$.}
\begin{tabular}{l|ccc}
\hline
Parameter & Analytical & Numerical & Difference \\
\hline
$K^*$ (rad) & $1.1204859\times10^{-2}$ & $1.272996\times10^{-2}$ & $13.6\%$ \\
$\epsilon$ (arcmin) & $0$ & $2.858$ & $(\ldots)$ \\
$T_u$ (y) & $2.225839$ & $2.22896$ & $0.14\%$ \\
$T_v$ (y) & $188.987571$ & $189.10854$ & $0.06\%$ \\
$T_w$ (y) & $346.236493$ & $348.49661$ & $0.65\%$ \\
\hline
\end{tabular}
\label{tab:comptitdr}
\end{table*}

\par Tables \ref{tab:frekdr} and \ref{tab:comptitdr} and Figure \ref{fig:titdr} summarize the result of the study of Titan's rotation with $\frac{C}{MR^2}=0.35$. Except for (c), the plots do not show any evident difference with Figure \ref{fig:tithr}, because the frequencies of the free librations are shifted just a little when $\frac{C}{MR^2}$ changes. In contrast, the behavior of the wobble $J$ (Figure \ref{fig:titdr}c) is very interesting, because this angle can be 10 times bigger than in the previous simulation. Table \ref{tab:etaxi} indicates that the most important terms in the solutions of $\xi_q$ and $\eta_q$, on which $J$ depends, are $\phi_w$ and $2\Phi_6$. In our cases, the periods of these terms are very close, so there might be a resonance between them, which could explain the amplitude of $J$. The matching on the frequencies of the free librations between the analytical and the numerical methods is still good, while a shift on Titan's mean obliquity still exists.

\section{Conclusion}

\par This paper offers a first study of Titan's rotation, where Titan is seen as a rigid body. We obtain a quasiperiodic decomposition of the forced solution, which can be split from the free solution in which Titan's obliquity plays an overwhelming role. Moreover, we find good matching between the frequencies of the free librations around the equilibrium, analytically and numerically evaluated, despite a model of circular orbit in the analytical study. However, we find a slight difference in the equilibrium obliquity. Finally, we cannot exclude a resonance between the proper mode $\Phi_6$ and Titan's wobble.

\par The next fly-bys of Cassini spacecraft should give us more information on Titan's gravitational field, so we should be able to make a more accurate study on its rotation, that could include direct perturbations on the other Saturnian satellites. These perturbations are supposed to be small (see for instance Henrard \cite{Henrard04}) and should be negligeable compared to the uncertainties we have on Titan's gravitational parameters. After that, the next step is to consider Titan as a multilayer non-rigid body and to study the consequences of its internal dissipation on the rotation.

\begin{acknowledgements}

\par The authors are indebted to J. Henrard and N. Rambaux for fruitful discussions. They also thank P. Tortora for having sent them an electronic copy of his DPS poster. This work has been supported by an FUNDP post-doctoral research grant.

\end{acknowledgements}

\begin{appendix}

\section{The coefficients $a_i$, $b_i$:}\label{sec:ab}

\begin{equation}
a_1=\frac{\sin^2I}{2}-\frac{1+\cos^2I}{4}=-\frac{1}{2}+3\gamma^2-3\gamma^4
\label{equ:a1}
\end{equation}

\begin{equation}
a_2=\frac{\sin2I}{2}=2\gamma\sqrt{1-\gamma^2}(1-2\gamma^2)
\label{equ:a2}
\end{equation}

\begin{equation}
a_3=\frac{\sin^2I}{8}=\frac{\gamma^2}{2}(1-\gamma^2)
\label{equ:a3}
\end{equation}

\begin{equation}
b_1=\frac{1+2\cos I+\cos^2I}{16}=\frac{1-2\gamma^2+\gamma^4}{4}
\label{equ:b1}
\end{equation}

\begin{equation}
b_2=\frac{2\sin I+\sin2I}{8}=\gamma(1-\gamma^2)\sqrt{1-\gamma^2}
\label{equ:b2}
\end{equation}

\begin{equation}
b_3=\frac{3}{8}\sin^2I=\frac{3}{2}\gamma^2(1-\gamma^2)
\label{equ:b3}
\end{equation}

\begin{equation}
b_4=\frac{2\sin I-\sin2I}{8}=\gamma^3\sqrt{1-\gamma^2}
\label{equ:b4}
\end{equation}

\begin{equation}
b_5=\frac{1-2\cos I+\cos^2I}{16}=\frac{\gamma^4}{4}
\label{equ:b5}
\end{equation}

\section{The coefficients $\gamma_{xx}$ and $\mu_{xx}$:}\label{sec:gamu}

\begin{equation}
\begin{split}
\gamma_{\sigma\sigma}=-2\delta_2(b_1(1+\cos K^*)^2+b_2\sin K^*(1+\cos K^*) \\
+b_3\sin^2K^*+b_4\sin K^*(1-\cos K^*)+b_5(1-\cos K^*)^2)
\end{split}
\label{equ:gsigsig}
\end{equation}

\begin{equation}
\begin{split}
\gamma_{\sigma\rho}=-\delta_2(b_2\sin K^*(1+\cos K^*)+b_3\sin^2K^* \\
+3b_4\sin K^*(1-\cos K^*)+4b_5(1-\cos K^*)^2)
\end{split}
\label{equ:gsigrho}
\end{equation}

\begin{equation}
\begin{split}
\gamma_{\rho\rho}=-\bigg(\delta_1\Big(\frac{a_2}{4}\sin 2K^*+4a_3\sin^2K^*\Big) \\
+\delta_2\Big(\frac{b_2}{2}\sin K^*(1+\cos K^*)+
2b_3\sin^2K^*+\\
\frac{9}{2}b_4\sin K^*(1-\cos K^*)+8b_5(1-\cos K^*)^2\Big)\bigg)
\end{split}
\label{equ:grhorho}
\end{equation}

\begin{equation}
\begin{split}
\gamma_{qq}=\frac{1}{2}\frac{\gamma_1+\gamma_2}{1-\gamma_1-\gamma_2} \\
-(\delta_1+\delta_2)\Big(\frac{\cos(K^*-I)}{4}+\frac{7}{16}\cos(2(K^*-I))+\frac{5}{16}\Big)
\end{split}
\label{equ:gqq}
\end{equation}

\begin{equation}
\begin{split}
\mu_{\sigma\sigma}=\frac{1}{2}+\frac{\delta_1}{P^{*2}}\Big((a_1+2a_3)(1-\cos K^*)(3\cos K^*-1) \\
+\frac{a_2}{2}\frac{\sin K^*}{(1+\cos K^*)^2}(6\cos^3K^*+4\cos^2K^*-5\cos K^*-2)\Big) \\
+\frac{\delta_2}{P^{*2}}\Big(b_1(3\cos K^*+1)(\cos K^*-1)+\frac{3}{2}b_2\frac{\sin K^*\cos 2K^*}{1+\cos K^*} \\
+b_3(1-\cos K^*)(3\cos K^*-1) \\
+\frac{b_4}{2}\frac{1-\cos K^*}{1+\cos K^*}\sin K^* (1+8\cos K^*+6\cos^2K^*)+3b_5\Big)
\end{split}
\label{equ:msigsig}
\end{equation}

\begin{equation}
\begin{split}
\mu_{\sigma\rho}=\frac{\delta_1}{P^{*2}}\Big((a_1+2a_3)(1-2\cos K^*)+ \\
\frac{a_2}{2}\frac{1+4\cos K^*-2\cos^2 K^*-4\cos^3 K^*}{\sin K^* (1+\cos K^*)}\Big)+ \\
\frac{\delta_2}{P^{*2}}\Big(2b_1\cos K^*-b_2\frac{4\cos^2K^*-\cos K^*-2}{2\sin K^*} \\
-b_3(2\cos K^*-1)+\frac{b_4}{2}\frac{\cos K^*}{\sin K^*}\frac{\cos K^* -1}{\cos K^* +1}(4\cos K^*+5) \\
-2b_5\frac{\sin^2K^*}{1+\cos K^*}\Big)
\end{split}
\label{equ:msigrho}
\end{equation}

\begin{equation}
\begin{split}
\mu_{\rho\rho}=-\frac{\delta_1}{P^{*2}}\Big(a_1+\frac{a_2}{2}\frac{3\cos K^*-2\cos^3K^*}{\sin^3K^*}+2a_3\Big)+ \\
\frac{\delta_2}{P^{*2}}\Big(b_1+\frac{b_2}{2}\frac{2\cos^3K^*-3\cos K^*-1}{\sin^3K^*} \\
-b_3-\frac{b_4}{2}\frac{1-3\cos K^*+2\cos^3K^*}{\sin^3K^*}+b_5\Big)
\end{split}
\label{equ:mrhorho}
\end{equation}

\begin{equation}
\mu_{qq}=\frac{1}{2}\frac{\gamma_1-\gamma_2}{1-\gamma_1+\gamma_2}
\label{equ:mqq}
\end{equation}

\end{appendix}

%\end{multicols}{2}


\begin{thebibliography}{}

\bibitem[1926]{Andoyer26} Andoyer H., 1926, \emph{Mécanique Céleste}, Gauthier-Villars, Paris

\bibitem[1990]{Bertotti90} Bertotti B. and Farinella P., 1990, \emph{Physics of the Earth and the Solar System}, Kluwer, A.P.

\bibitem[2003]{Bouquillon03} Bouquillon S., Kinoshita H. and Souchay J., 2003, Celes. Mech. Dyn. Astr., 86, 29

\bibitem[1961]{Brouwer61} Brouwer D. and Clemence G.M., 1961, \emph{Methods of Celestial Mechanics}, Academic Press, New-York

\bibitem[1989]{Campbell89} Campbell J.K. and Anderson J.D., 1989, AJ, 97, 1485

\bibitem[1987]{Carpino87} Carpino M., Milani A. and Nobili A.M., 1987, A\&A, 181, 182

\bibitem[1998]{Champenois98} Champenois S., 1998, \emph{Dynamique de la résonance entre Mimas et Téthys, premier et troisième satellites de Saturne}, Ph.D thesis, Observatoire de Paris

\bibitem[1967]{Deprit67} Deprit A., 1967, American Journal of Physics, 35, 424

\bibitem[2004]{DHoedt04} D'Hoedt S. and Lemaître A., 2004, Cel. Mech. Dyn. Astr., 89, 267

\bibitem[2004]{Henrard04} Henrard J. and Schwanen G., 2004, Cel. Mech. Dyn. Astr., 89, 181

\bibitem[2005a]{Henrard05i} Henrard J., 2005a, Icarus, 178, 144

\bibitem[2005b]{Henrard05c} Henrard J., 2005b, Cel. Mech. Dyn. Astr., 91, 131

\bibitem[2005c]{Henrard052} Henrard J., 2005c, Cel. Mech. Dyn. Astr., 93, 101

\bibitem[2006]{Lainey06} Lainey V., Duriez L. and Vienne A., 2006, A\&A, 456, 783

\bibitem[1988]{Laskar88} Laskar J., 1988, A\&A, 198, 341

\bibitem[1992]{Laskar92} Laskar J., Froeschlé Cl. and Celletti A., 1992, Physica D, 56, 253

\bibitem[2004]{Laskar04} Laskar J., 2004, \emph{Frequency analysis, quasiperiodic decompositions, and Nyquist limit} in \emph{Journées scientifiques 2003 de l'Institut de Mécanique Céleste et de Calcul des \'Ephémérides}, Notes Scientifiques et Techniques de l'Institut de Mécanique Céleste, S081, 29

\bibitem[1993]{Lemmon93} Lemmon M.T., Karkoschka E. and Tomasko M., 1993, Icarus, 103, 329

\bibitem[1995a]{Lemmon95} Lemmon M.T., Karkoschka E. and Tomasko M., 1995, Icarus, 113, 27

\bibitem[2004]{Rambaux04} Rambaux N. and Bois E., 2004, A\&A, 413, 381

\bibitem[2005]{Rambaux05} Rambaux N. and Henrard J., 2005, \emph{The rotation of the Galilean satellites}, in \emph{The rotation of celestial bodies}, ed. A. Lemaître, Presses Universitaires de Namur, Namur

\bibitem[2004]{Richardson04} Richardson J., Lorenz R.D. and McEwen A., 2004, Icarus, 170, 113

\bibitem[2002]{Seidelmann02} Seidelmann P.K., Abalakin V.K., Bursa M. et al., 2002, Cel. Mech. Dyn. Astr., 82, 83

\bibitem[2006]{Tortora06} Tortora P., Armstrong J.W., Asmar S.W. et al., 2006, \emph{The determination of Titan's gravity field with Cassini}, DPS meeting 38, 56.01

\bibitem[1995]{Vienne95} Vienne A. and Duriez L., 1995, A\&A, 297, 588

\end{thebibliography}
\end{document}